\documentclass[letterpaper]{article} 
\usepackage{aaai2026}  
\usepackage{times}  
\usepackage{helvet}  
\usepackage{courier}  
\usepackage[hyphens]{url}  
\usepackage{graphicx} 
\usepackage{natbib}  
\usepackage{caption} 
\usepackage{algorithm}
\usepackage{algorithmic}

\usepackage{newfloat}
\usepackage{listings}
\DeclareCaptionStyle{ruled}{labelfont=normalfont,labelsep=colon,strut=off} 
\floatstyle{ruled}
\newfloat{listing}{tb}{lst}{}
\floatname{listing}{Listing}
\usepackage{amsfonts}
\usepackage{amsmath}
\usepackage{multirow}
\usepackage{multicol}

\title{RL-U$^2$Net: A Dual-Branch UNet with Reinforcement Learning-Assisted Multimodal Feature Fusion for Accurate 3D Whole-Heart Segmentation}

\author {
    Jierui Qu\textsuperscript{\rm 1},
    Jianchun Zhao\textsuperscript{\rm 2}\thanks{Corresponding author}
}
\affiliations {
    \textsuperscript{\rm 1}College of Design and Engineering, National University of Singapore, 117575, Singapore\\
    \textsuperscript{\rm 2}School of Electrical Engineering, Xi'an Jiaotong University, Xi'an 710049, China\\
    zhao\_jianchun@stu.xjtu.edu.cn
}

\usepackage{bibentry}

\begin{document}

\maketitle

\begin{abstract}
Accurate whole-heart segmentation is a critical component in the precise diagnosis and interventional planning of cardiovascular diseases. Integrating complementary information from modalities such as computed tomography (CT) and magnetic resonance imaging (MRI) can significantly enhance segmentation accuracy and robustness. However, existing multi-modal segmentation methods face several limitations: severe spatial inconsistency between modalities hinders effective feature fusion; fusion strategies are often static and lack adaptability; and the processes of feature alignment and segmentation are decoupled and inefficient. To address these challenges, we propose a dual-branch U-Net architecture enhanced by reinforcement learning for feature alignment, termed RL-U$^2$Net, designed for precise and efficient multi-modal 3D whole-heart segmentation. The model employs a dual-branch U-shaped network to process CT and MRI patches in parallel, and introduces a novel RL-XAlign module between the encoders. The module employs a cross‑modal attention mechanism to capture semantic correspondences between modalities and a reinforcement‑learning agent learns an optimal rotation strategy that consistently aligns anatomical pose and texture features. The aligned features are then reconstructed through their respective decoders. Finally, an ensemble‑learning–based decision module integrates the predictions from individual patches to produce the final segmentation result. Experimental results on the publicly available MM-WHS 2017 dataset demonstrate that the proposed RL-U$^2$Net outperforms existing state-of-the-art methods, achieving Dice coefficients of 93.1\% on CT and 87.0\% on MRI, thereby validating the effectiveness and superiority of the proposed approach.
\end{abstract}

\begin{links}
    \link{Code}{https://github.com/TantalumKevin/RL-U2NET}
\end{links}

\section{Introduction}
Cardiovascular disease (CVD) represents a leading cause of mortality worldwide. Accurate three-dimensional whole-heart segmentation is essential for quantitative lesion assessment and clinical decision-making. While computed tomography (CT) and magnetic resonance imaging (MRI) serve as primary diagnostic tools, single-modality approaches suffer from inherent limitations in contrast, spatial resolution, and imaging artifacts that compromise comprehensive cardiac characterization. Effective multi-modal feature fusion methods are therefore critical for enhancing segmentation accuracy and robustness in clinical applications~\cite{valsangiacomobuechel2012imaging,puyol-anton2022multimodal}.

\begin{figure}[t]
\centering
\includegraphics[width=0.99\columnwidth]{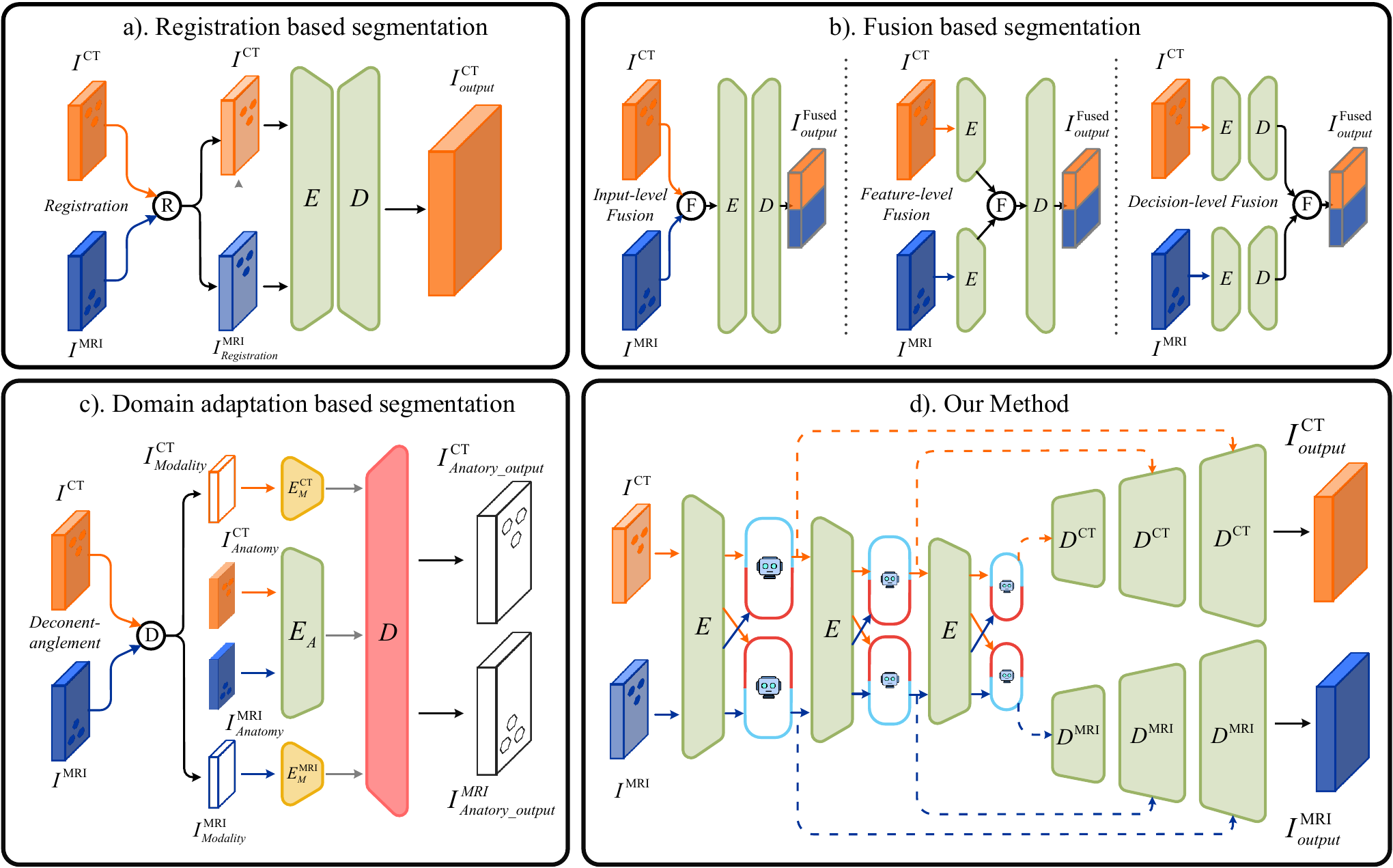}
\caption{Paradigm comparisons between the existing multimodal medical image segmentation methods and our method}
\label{fig:introduction}
\end{figure}

Deep learning has significantly advanced CVD diagnosis through medical imaging analysis. U-shaped architectures have become the dominant paradigm for medical segmentation due to their multi-scale feature aggregation and skip connections~\cite{ronneberger2015unet,jin2020raunet}. However, traditional 3D CNNs suffer from limited receptive fields, hindering long-range dependency modeling~\cite{cicek20163d}. While Transformers' self-attention mechanisms capture global correlations and achieve notable progress in visual segmentation~\cite{carion2020endtoend,strudel2021segmenter,cao2022swinunet}, they exhibit weaker fine-grained local feature representation and higher computational costs. To address these limitations, hybrid CNN-Transformer frameworks have emerged~\cite{chen2021transunet,wang2022mixed,chen2021transunet}. This integration of CNN's local discriminative capabilities with Transformer's global dependency modeling has proven effective for improving cardiac segmentation accuracy.

Cardiac imaging is challenged by the heart's non-rigid dynamics, often necessitating multi-modal approaches as single modalities provide incomplete information~\cite{freed2016mr}. Consequently, multi-modal cardiac segmentation has drawn significant interest, typically employing registration, fusion, or domain adaptation to mitigate inter-modal discrepancies~\cite{li2023multimodality} (Figure~\ref{fig:introduction}). However, existing methods suffer from three critical limitations. First, substantial spatial misalignment, induced by cardiac and respiratory motion, renders traditional image-level registration inadequate for precise correspondence. Second, most fusion strategies rely on static concatenation or simple weighting, lacking deep understanding of inter-modal semantic relationships and limiting feature expressiveness. Third, many approaches decouple feature alignment from the segmentation objective, precluding end-to-end optimization and compromising overall performance.

To address these challenges, we propose RL-U$^2$Net, a reinforcement learning-assisted dual-branch network for multimodal feature alignment. The architecture employs parallel encoders to process CT and MRI patches while preserving modality-specific characteristics. The core RL-XAlign module, inserted between encoders, first establishes semantic correspondences via cross-modal attention, then leverages a reinforcement learning agent to learn optimal spatial alignment strategies for cross-modal features. Aligned features undergo modality-specific reconstruction through dedicated decoders. To ensure training stability, an adaptive gradient weight distributor(AGWD) dynamically balances inter-modal gradient differences, while an ensemble-based decision module integrates patch-level predictions for final segmentation. This network effectively unifies feature alignment, adaptive fusion, and end-to-end optimization. Experiments on MM-WHS 2017 demonstrate state-of-the-art performance, validating our approach's effectiveness. The main contributions are:

\begin{itemize}
\item This paper introduces for the first time a cross-modal feature alignment module assisted by reinforcement learning. It extracts semantic correspondences through a cross-modal attention mechanism and utilizes a reinforcement learning agent to dynamically learn the optimal three-dimensional rotation strategy, effectively solving the problem of spatial inconsistency between multimodal images.
\item Design an AGWD that dynamically adjusts the gradient weights of the two modalities during the training phase to maintain stability and balance in the optimization process and promote collaborative learning of dual-modality features.
\item Construct a dual-branch U-Net structure to process CT and MRI modalities separately, and introduce a decision module based on ensemble learning to fuse patch-level prediction results, thereby improving the accuracy and robustness of multi-modal whole-heart segmentation.
\end{itemize}

\begin{figure*}[t]
\centering
\includegraphics[width=0.99\textwidth]{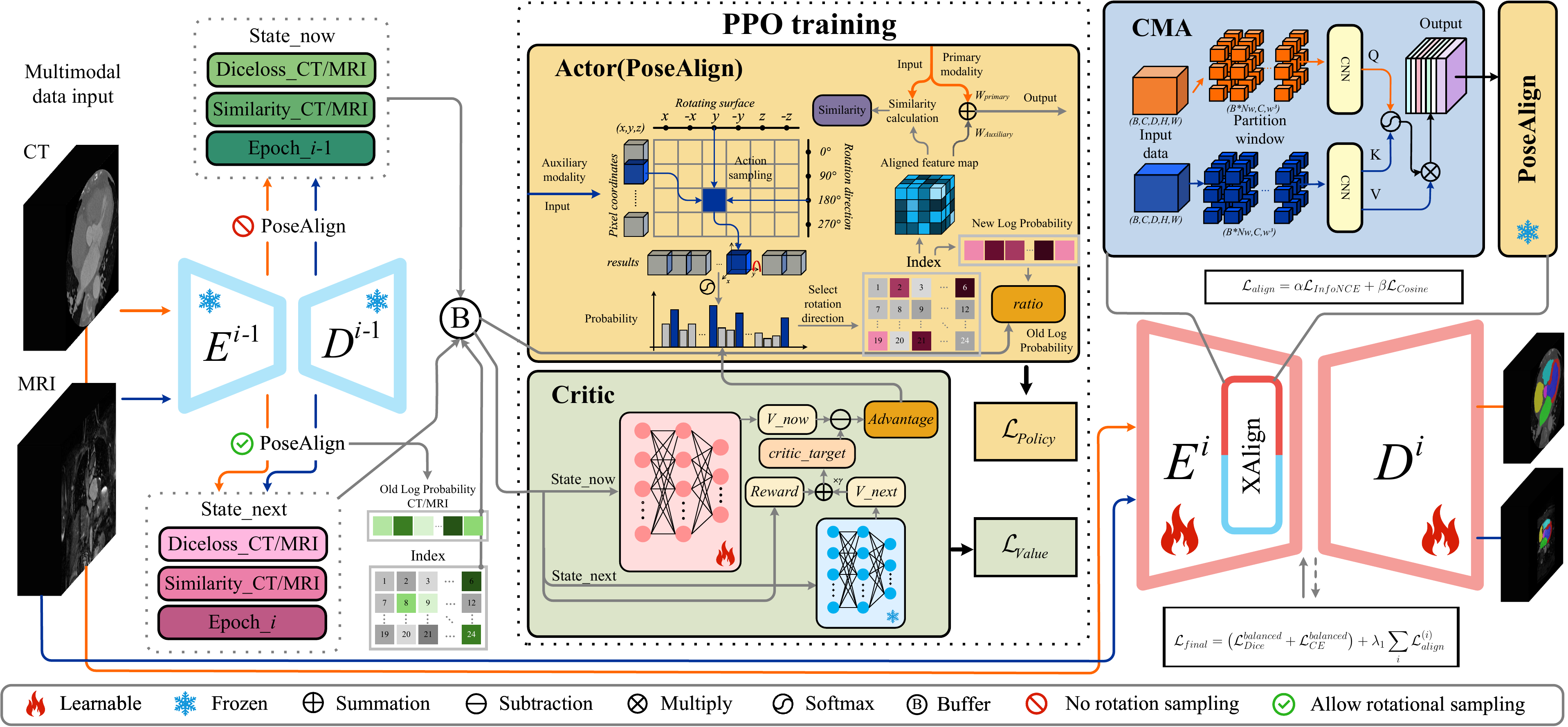}
\caption{Overall framework of PPO training in RL-XAlign module. The framework demonstrates how CMA establishes semantic correspondences while PPO's Actor-Critic architecture learns optimal spatial alignment strategies through iterative state-action optimization for cross-modal feature fusion.}
\label{fig:PPO-training}
\end{figure*}

\section{Related Work}
\subsection{U-Net for 3D Medical Image Segmentation}
To overcome 3D CNN limitations in long-range dependency modeling, UNETR integrates Transformer encoders with U-shaped decoders~\cite{hatamizadeh2022unetr}, while Swin-UNETR employs hierarchical shifted window attention with self-supervised pre-training~\cite{tang2022selfsupervised,hatamizadeh2021swin}. Recent advances include axial global attention (GASA-UNet)~\cite{sun2024gasaunet}, Mamba-based state space models (EM-Net)~\cite{chang2024emnet}, and multi-scale convolution-attention fusion~\cite{pan2025multiscale}, reflecting efforts to balance computational efficiency with global context modeling. However, these methods primarily target single-modal scenarios and lack explicit handling of spatial inconsistencies and semantic correspondences in multimodal data, making feature alignment and fusion critical performance bottlenecks.

\subsection{Multimodal Cardiac Segmentation}
Multimodal cardiac segmentation has attracted considerable research interest~\cite{zhuang2020myocardial}. Existing approaches fall into three categories: registration-based methods that spatially align modalities before segmentation~\cite{luo2022mathcal,zhuang2018multivariate,luo2020mvmmregnet}; fusion-based methods that exploit complementary CT-MRI characteristics through input-level~\cite{yu2020dual,zhang2020fully}, feature-level~\cite{zhao2020stacked,li2022myocardial}, or decision-level fusion~\cite{rokach2010ensemblebased} with attention mechanisms; and domain adaptation methods using adversarial learning or style transfer for cross-domain generalization~\cite{pei2021disentangle,koehler2021unsupervised,wang2022cycmis}. However, these methods typically treat registration and fusion as preprocessing steps~\cite{li2023myops}, making unified end-to-end optimization of features alignment, fusion, and segmentation a persistent challenge.

\subsection{Reinforcement Learning and PPO Algorithms for Vision Tasks}
Reinforcement learning (RL) learns optimal policies through agent-environment interactions to maximize long-term rewards~\cite{kaelbling1996reinforcement}. Deep RL employs neural networks for policy and value function modeling, enabling high-dimensional applications~\cite{arulkumaran2017brief}. In vision tasks, RL's iterative "perception-correction-feedback" process proves particularly effective for complex organ segmentation with ambiguous boundaries. Recent pixel-level RL methods have improved multi-organ boundary accuracy~\cite{liu2025pixel}, while RL agents excel in image navigation, keypoint localization, and contour refinement~\cite{alansary2018automatic,ghesu2016artificial,liao2020iterativelyrefined}. Proximal Policy Optimization (PPO) achieves optimal balance among sample efficiency, stability, and implementation complexity through clipped surrogate objectives that constrain policy updates~\cite{schulman2017proximal}. With the development of multimodal large models, PPO has been widely used for cross-modal policy optimization~\cite{wan2025srpo,huang2025highresolution,he2016deep}. Inspired by this, this paper innovatively introduces the PPO algorithm into medical image segmentation, utilizing reinforcement learning to assist in multimodal feature alignment.

\section{Method}

\subsection{Overview}
The main network of RL-U$^2$Net consists of a shared encoder based on Swin Transformer, an RL-XAlign cross-modal alignment module, a Res-Fusion fusion module, and a dual-branch ResU-Net decoder. Due to space limitations, the detailed overview is included in the supplementary materials. The following sections will detail the design principles and implementation mechanisms of each core module.

\subsection{Reinforcement Learning based RL-XAlign module}

The RL-XAlign module adopts a reinforcement learning framework, modeling cross-modal feature alignment as a sequential decision-making process, as shown in Figure~\ref{fig:PPO-training}. For encoder layer $i$ with CT and MRI feature maps $F_{\rm{CT}}^{i-1}$ and $F_{\rm{MRI}}^{i-1}$, the module first employs cross-modal attention (CMA) to capture semantic correspondences and construct preliminary cross-modal representations. The PoseAlign component then treats current features as environment states, where an RL agent selects optimal actions from 24 predefined 3D rotations via policy networks for precise spatial alignment. Training utilizes Proximal Policy Optimization (PPO) with Actor-Critic architecture to simultaneously optimize policy and value networks. Through iterative optimization, the module adaptively learns optimal alignment strategies, outputting spatially consistent and semantically aligned features $F_{\rm{CT}}^{i}$ and $F_{\rm{MRI}}^{i}$ for subsequent segmentation tasks.

\subsubsection{CMA Module}
The CMA module is the primary component of the RL-XAlign module, responsible for establishing semantic correspondences between different modalities. Considering the high-dimensional characteristics of 3D medical images and computational efficiency requirements, this study designed a CMA module based on segmentation windows. For the input CT and MRI feature maps $F_{\rm{CT}}^{(i-1)},F_{\rm{MRI}}^{(i-1)}\in\mathbb{R}^{B\times C\times D\times H\times W}$, with CT as the primary modality and MRI as the auxiliary modality, the CMA module first divides the 3D feature maps into non-overlapping cubic windows of size $w\times w\times w$, converting global attention calculation into local attention calculation within the window, thereby effectively reducing computational complexity.

Within each window, CMA performs cross-modal attention calculations. The query, key, and value matrices are linearly projected through independent 1D convolution layers:

\begin{equation}
    Q=\mathrm{Conv}1\mathrm{D}_{q}(W(F_{\rm{CT}}^{(i-1)}))
\end{equation}

\begin{equation}
    K=\mathrm{Conv}1\mathrm{D}_{k}(W(F_{\rm{MRI}}^{(i-1)}))
\end{equation}

\begin{equation}
    V=\mathrm{Conv}1\mathrm{D}_{v}(W(F_{\rm{MRI}}^{(i-1)}))
\end{equation}

Where $W(\cdot)$ denotes the window segmentation operation. The cross-modal attention calculation formula within the window is:

\begin{equation}
    \mathrm{Attention}(Q,K,V)=\mathrm{softmax}\left(\frac{Q^TK}{\sqrt{d}}\right)V^T
\end{equation}

Finally, the original feature map size is restored through output projection layer and window inverse transformation operations:

\begin{equation}
    F_{\mathrm{cross}}^{i}=W^{-1}(\mathrm{Conv}1\mathrm{D}_o(\mathrm{Attention}(Q,K,V)))
\end{equation}

This mechanism enables bidirectional information exchange between CT and MRI features, generating feature representations that fuse cross-modal semantic information and provide feature inputs rich in complementary information for the subsequent pose alignment stage.

\subsubsection{PoseAlign Module}
After obtaining cross-modal semantic feature representations, the PoseAlign module is responsible for solving spatial inconsistencies between multimodal modalities. This module abstracts the 3D spatial alignment problem into a discrete rotation transformation selection process, achieving precise feature pose correction through a predefined set of rotation actions.

The core design of the PoseAlign module is based on the theory of cube rotational symmetry, constructing a complete action space comprising 24 rotational transformations~\cite{worrall2018cubenet}. These 24 rotations encompass all possible orientations of a cube in 3D space. The specific generation process is achieved by combining the six face orientations ($\pm x,\pm y,\pm z$ axis directions) with four rotation angles (0°, 90°, 180°, 270°) for each face. The mathematical representation of the rotation matrix is:

\begin{equation}
    \mathcal{R}=\{R_{1},R_{2},\ldots,R_{24}\}\subset SO(3)
\end{equation}

Each rotation matrix $R_i\in\mathbb{R}^{3\times3}$ corresponds to a unique 3D rotation transformation. To improve the learning efficiency of reinforcement learning, this module randomly samples $K_{sel}$ from 24 rotation transformations according to the learning weights of each direction during each forward propagation, and uses the mean of the rotated feature map as the output $F_{aligned}$

The RL-XAlign module ultimately fuses the aligned auxiliary modal feature maps into the main mode according to certain weights:

\begin{equation}
    F_{O_{\mathrm{CT}}}^{i}=\lambda\cdot F_{\rm{aligned}}^{i}+(1-\lambda)\cdot F_{\rm{CT}}^{i}
\end{equation}

Among them, $\lambda$ is the preset fusion weight, which ensures that the main modal features dominate, while the aligned auxiliary modal features supplement cross-modal information with lower weights.

\subsubsection{Reinforcement Learning Training Strategies}
The training process based on the PPO algorithm is the core driving mechanism of the RL-XAlign module, which models cross-modal feature alignment as a Markov decision process and implements collaborative learning of policy optimization and value assessment through the Actor-Critic architecture~\cite{yao2024cmragent}. This training strategy uses experience replay and policy pruning mechanisms to ensure the stability and convergence of the training process.
The state space of the reinforcement learning environment is designed as a three-dimensional vector representation:

\begin{equation}
    s_t=[(1-\mathcal{L}_{Dice}),S_{similarity},epoch]
\end{equation}

The dice coefficient reflects the current segmentation quality, the similarity metric measures the degree of feature alignment between modalities, and the training progress factor provides temporal prior information. The action space corresponds to 24 predefined rotation transformations, and the agent needs to learn to select the optimal rotation strategy in a given state to maximize the cumulative reward. The reward function is designed as a weighted combination of segmentation performance and feature alignment quality:

\begin{equation}
    r_t=(1-\mathcal{L}_{Dice})+S_{similarity}
\end{equation}

Where $\mathcal{L}_{Dice}$ is the Dice loss and $S_{similarity}$ is the cross-modal feature similarity, the design allows the intelligences to optimize both segmentation accuracy and feature alignment quality.

The PPO training process adopts the Actor-Critic framework, in which the Actor network consists of learnable parameters $\theta\in\mathbb{R}^{24}$ in the PoseAlign module, which generates an action probability distribution $\pi_\theta(a|s)$ through a softmax operation. The Critic network is a simple multi-layer perceptron structure that inputs a state vector and outputs a value estimate $V_{\phi}(s)$.

The training process is divided into two stages: experience collection and strategy update. In the experience collection stage, the intelligent agent interacts with the environment to generate trajectory samples $(s_t, a_t, r_t, s_{t+1})$ and stores them in the experience buffer. In the strategy update stage, parameter optimization is performed using the pruning objective function of PPO.

The advantage function estimate is calculated using the time difference method:

\begin{equation}
    A_t=r_t+\gamma V_\phi(s_{t+1})-V_\phi(s_t)
\end{equation}

Standardized processing is performed to reduce variance. The policy loss uses the pruning objective function of the PPO algorithm:

\begin{align}
    \notag \mathcal{L}_{Policy} = -\mathbb{E} \left[ A_t \cdot \min \left(\frac{\pi_\theta(a_t|s_t)} {\pi_{\theta_{old}}(a_t|s_t)},\right. \right.\\
    \left. \left. \mathrm{clip}_{\epsilon} \left( \frac {\pi_\theta(a_t|s_t)} {\pi_{\theta_{old}}(a_t|s_t)}\right) \right) \right]
    \label{equ:l-policy}
\end{align}

Where $\mathbb{E}(\cdot)$ denotes the expectation operator (i.e., the mean value), the function $\mathrm{clip}_{\epsilon}(\cdot)$ ensures that the result remains within the range  $\left[1-\epsilon,1+\epsilon\right]$, and $\epsilon$ is the clipping parameter, which limits the policy ratio to a reasonable range. The value function loss is expressed as the mean square error:

\begin{equation}
    \mathcal{L}_{Value}=\mathbb{E}\left[\left(V_\phi(s_t)-V_t^{target}\right)^2\right]
    \label{equ:l-value}
\end{equation}

The target value is:

\begin{equation}
    V_{t}^{target}=r_{t}+\gamma V_{\phi}(s_{t+1})
\end{equation}

The pseudocode of the PPO training is as follows.

The training algorithm employs a multi-round mini-batch update strategy, with each training cycle comprising multiple experience sampling and parameter updates. Specifically, for each RL-XAlign module, the algorithm first collects $N$ trajectory samples, then performs $K$ rounds of mini-batch updates, with each round randomly sampling a batch size of $M$ experiences for gradient descent. This design achieves a good balance between sample efficiency and training stability, enabling the agent to quickly converge to the optimal policy in complex multi-modal alignment tasks.

\subsection{Decision-making Module Based on Ensemble Learning}
\begin{figure}[t]
\centering
\includegraphics[width=0.99\columnwidth]{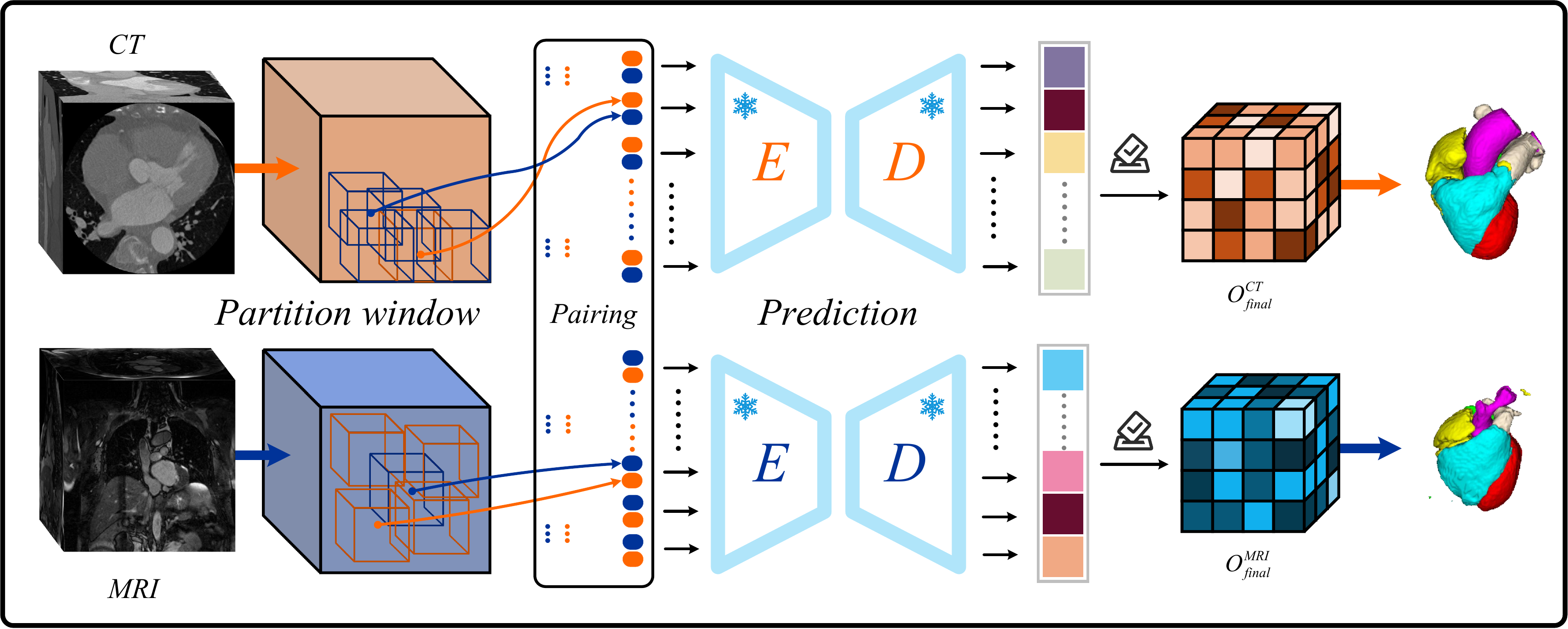}
\caption{Schematic of the decision-making module based on ensemble learning.}
\label{fig:ensemble-decision-module}
\end{figure}

The final prediction stage of RL-U$^2$Net adopts a decision strategy based on ensemble learning, achieving high-precision whole-heart segmentation through a sliding window inference mechanism and multi-level voting fusion. This module decomposes large-sized 3D medical images into overlapping local windows, with each window acting as an independent weak learner for prediction. The final segmentation result is generated through a weighted voting strategy to achieve a globally consistent segmentation outcome, as shown in the Figure~\ref{fig:ensemble-decision-module}.

The ensemble decision module adopts a sliding window inference framework. First, the input image $I\in\mathbb{R}^{H\times W\times D}$ is densely sampled according to the preset region of interest (ROI) size to generate a series of overlapping 3D windows. The scanning interval is adaptively calculated using an overlap ratio parameter to ensure appropriate overlap between adjacent windows and reduce edge artifacts. To further enhance ensemble performance, the module supports cross-slice ensemble mode, which establishes correspondences between different modalities through a spatial mapping mechanism. This mechanism calculates scaling transformation factors based on the spatial resolution differences between CT and MRI images, then constructs spatially transformed window regions, retaining only valid window pairs with spatial overlap for subsequent processing.

To ensure smooth blending of overlapping areas, the module calculates a Gaussian-based importance weight map for each window, giving higher confidence to the center area of the window and gradually decreasing the weight toward the boundary areas:

\begin{equation}
    w(x,y,z)=\exp\left(-\frac{(x-x_c)^2+(y-y_c)^2+(z-z_c)^2}{2\sigma^2}\right)
\end{equation}

where $(x_c, y_c, z_c)$ are the window center coordinates, and $\sigma$ is the scale parameter. During inference, each window independently generates segmentation predictions $P_{\rm{MRI}}^{(j)}$ and $P_{\rm{CT}}^{(j)}$ for CT and MRI using the RL-U$^2$Net model, which are then accumulated into the global output buffer based on their spatial positions:

\begin{equation}
    O_{final}^{\rm{CT}}(x,y,z)=\frac{\sum_{j}w_{j}(x,y,z)\cdot P_{\rm{CT}}^{(j)}(x,y,z)}{\sum_{j}w_{j}(x,y,z)}
\end{equation}

\begin{equation}
    O_{final}^{\rm{MRI}}(x,y,z)=\frac{\sum_{j}w_{j}(x,y,z)\cdot P_{\rm{MRI}}^{(j)}(x,y,z)}{\sum_{j}w_{j}(x,y,z)}
\end{equation}

This weighted averaging mechanism implements a soft voting strategy, which has good numerical stability and boundary continuity. Multiple predictions in overlapping regions are automatically constrained for consistency through weight normalization, effectively eliminating discontinuities at window boundaries and providing stable and reliable prediction results for clinical applications.

\subsection{Loss Function}
\subsubsection{Multi-task Loss Function System}
RL-U$^2$Net constructs a complete loss function system covering segmentation supervision, cross-modal alignment, and reinforcement learning training. Through the collaborative optimization of multiple subtasks, it achieves joint improvement in feature alignment and segmentation performance.

The segmentation supervision loss employs a combination strategy of Dice loss $\mathcal{L}_{Dice}$ and cross-entropy loss $\mathcal{L}_{CE}$, which ensures overall segmentation performance while enhancing the accuracy of detailed boundaries. The cross-modal alignment loss specifically optimizes the feature alignment quality in the RL-XAlign module, consisting of InfoNCE loss $\mathcal{L}_{InfoNCE}$ and cosine embedding loss$\mathcal{L}_{Cosine}$. InfoNCE loss~\cite{oord2018representation} adopts contrastive learning ideas, learning discriminative feature representations by maximizing the similarity of positive samples and minimizing the similarity of negative samples. The cosine embedding loss~\cite{payer2019segmenting} directly constrains the cosine similarity of aligned features.And the total loss for cross-modal alignment is:

\begin{equation}
    \mathcal{L}_{align}=\alpha\mathcal{L}_{InfoNCE}+\beta\mathcal{L}_{Cosine}
\end{equation}

Where $\alpha$ and $\beta$ are equilibrium weights.

The reinforcement learning loss function is responsible for optimizing the policy network and value network in the PoseAlign module. The policy loss $\mathcal{L}_{Policy}$ and value loss $\mathcal{L}_{Value}$ are shown in equations (\ref{equ:l-policy}) and (\ref{equ:l-value}).

\subsubsection{Adaptive Gradient Weight Distributor(AGWD)}

CT and MRI modalities often exhibit different levels of learning difficulty due to differences in imaging mechanisms, contrast characteristics, and anatomical structure representation. Traditional fixed-weight loss functions cannot adapt to dynamic changes in learning states, often leading to overfitting in one modality and underfitting in another. To address this, this paper proposes an AGWD that dynamically monitors the learning progress of each modality and adjusts loss weights in real-time to achieve adaptive balance in multi-modal learning. Weight calculations use a hyperbolic tangent function for smooth adjustment:

\begin{equation}
    w=\tanh\left(\gamma\cdot(\mathcal{L}_{slow}-\mathcal{L}_{fast}-\delta)\right)
\end{equation}

Where $\mathcal{L}_{slow}$ and $\mathcal{L}_{fast}$ represent the loss values of the slow convergence mode and fast convergence mode, respectively, $\gamma$ is the temperature parameter controlling the sensitivity of weight adjustment, and $\delta$ is the baseline offset providing stability assurance. The selection of the hyperbolic tangent function ensures that the weight values vary within the $(-1,1)$ interval, avoiding training instability caused by extreme weights. Based on the calculated weight factors, the balanced loss is reconstructed using an adaptive weighting strategy:

\begin{equation}
    \mathcal{L}^{balanced}=\mathcal{L}^{fast}(1-w)+\mathcal{L}^{slow}(1+w)
\end{equation}

This weighting strategy is applied to both Dice loss and cross-entropy loss. After integration with the multi-task loss function system, the final loss function is:

\begin{equation}
    \mathcal{L}_{final}=(\mathcal{L}_{Dice}^{balanced}+\mathcal{L}_{CE}^{balanced})+\lambda_{1}\sum_{i}\mathcal{L}_{align}^{(i)}
\end{equation}

$\mathcal{L}_{align}^{(i)}$ is the alignment loss for layer $i$, and $\lambda_1$ is the weight balancing parameter. This design achieves adaptive balancing of segmentation supervision loss and collaborative optimization of cross-modal alignment loss, ensuring that the entire network maintains a stable and efficient training process in complex multimodal learning tasks.

\section{Results}
\subsection{Datasets and Pre-Processings}
The dataset used in this study was obtained from the MM-WHS Challenge 2017~\cite{zhuang2019evaluation}, which includes 60 sets of cardiac CT and MRI image data. Following the data partitioning strategy adopted in previous studies~\cite{cui2023improved,cui2025p2tc}, the 40 sets of data were used as the training set, while the 20 sets of labeled data were randomly divided into 15 sets for testing and 5 sets for validation. The rest of the information of dataset is included in the supplementary materials.

\subsection{Results on MM-WHS 2017 Challenge Dataset}
\begin{table*}[t]
\centering
\begin{tabular}{cccccccccc}
\hline
\multirow{2}*{Method} & \multicolumn{8}{c}{Dice$\uparrow$} & \multirow{2}*{HD95(mm)$\downarrow$} \\
\cline{2-9}
~ & Myo & LA & LV & RA & RV & AA & PA & Average & ~ \\
\hline
3D U-Net~\cite{cicek20163d} & 0.894 & 0.909 & 0.917 & 0.869 & 0.891 & 0.933 & 0.883 & 0.899 & 22.988 \\
ConResNet~\cite{lee20223d} & 0.918 & 0.929 & 0.928 & 0.883 & 0.914 & 0.949 & 0.852 & 0.910 & 26.652 \\
nnformer~\cite{zhou2023nnformer} & 0.866 & 0.916 & 0.923 & 0.899 & 0.917 & 0.935 & 0.873 & 0.904 & 12.174 \\
D-Former~\cite{wu2023dformer} & 0.860 & 0.892 & 0.918 & 0.903 & 0.920 & 0.937 & 0.886 & 0.902 & 14.760 \\
SwinUNETR & \multirow{2}*{0.875} & \multirow{2}*{0.926} & \multirow{2}*{0.924} & \multirow{2}*{0.891} & \multirow{2}*{0.922} & \multirow{2}*{0.931} & \multirow{2}*{0.885} & \multirow{2}*{0.908} & \multirow{2}*{17.664} \\
\cite{hatamizadeh2021swin}\\
UNETR++~\cite{shaker2024unetr} & 0.883 & 0.881 & 0.924 & 0.899 & 0.893 & 0.934 & 0.860 & 0.896 & 14.850 \\
Cascaded U-Net & \multirow{2}*{0.899} & \multirow{2}*{0.921} & \multirow{2}*{0.927} & \multirow{2}*{0.905} & \multirow{2}*{0.909} & \multirow{2}*{0.946} & \multirow{2}*{0.889} & \multirow{2}*{0.914} & \multirow{2}*{14.163} \\
\cite{salgado-garcia2024efficient}\\
HRMedSeg~\cite{xu2025hrmedseg} & 0.910 & 0.924 & 0.937 & 0.913 & 0.920 & 0.951 & 0.892 & 0.921 & 12.255 \\
P2TC~\cite{cui2025p2tc} & 0.907 & 0.930 & 0.936 & 0.894 & 0.918 & 0.953 & 0.889 & 0.918 & 21.417 \\
\hline
RL-U$^2$Net(Ours) & \textbf{0.927} & \textbf{0.947} & \textbf{0.938} & \textbf{0.922} & \textbf{0.933} & \textbf{0.959} & \textbf{0.894} & \textbf{0.931} & \textbf{11.471}\\
\hline
\end{tabular}
\caption{Performance comparison of RL-U$^2$Net and the SOTA segmentation methods on the MM-WHS 2017 CT dataset.}
\label{table:ct}
\end{table*}

To comprehensively evaluate the segmentation performance of RL-U$^2$Net, we systematically compared it with nine state-of-the-art segmentation models on the MM-WHS 2017 dataset, as shown in the Table~\ref{table:ct}. To ensure fairness in the comparison, all comparison methods were re-experimented locally using the same dataset and evaluation metrics, with Dice coefficient and Hausdorff distance as the primary evaluation metrics. Training strategy and implementation details is included in the supplementary materials.

In CT image segmentation tasks, RL-U$^2$Net demonstrated outstanding overall performance, with an average Dice coefficient of 93.1\%, which is 1 percentage point higher than the second-best method, HRMedSeg~\cite{xu2025hrmedseg}, with 92.1\%. In terms of boundary accuracy, RL-U$^2$Net had an average Hausdorff distance of 11.471 mm, which was significantly better than all other methods. Results of MRI image segmentation tasks is included in the supplementary materials.

\begin{figure}[t]
\centering
\includegraphics[width=0.99\columnwidth]{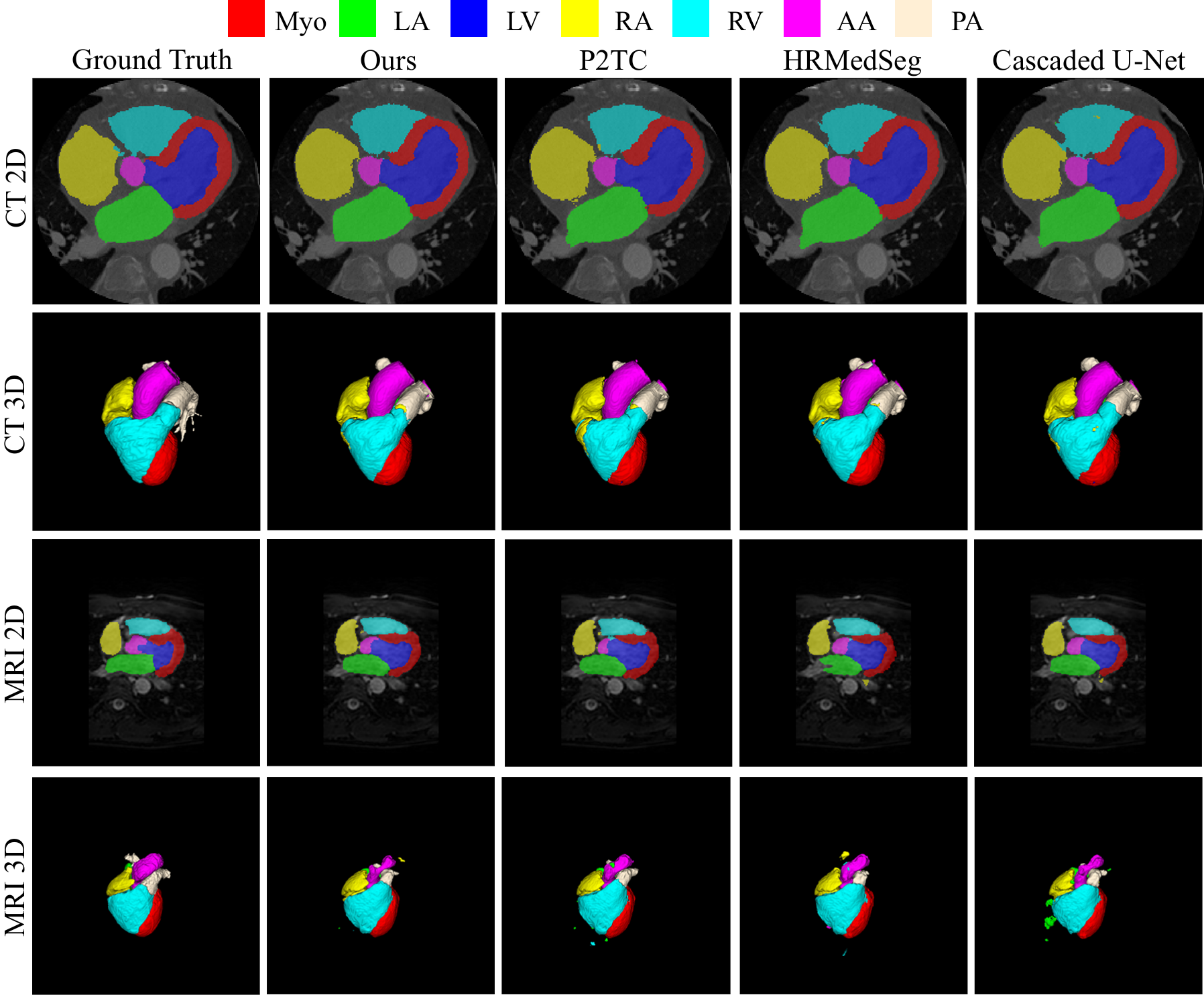}
\caption{Visualization of methods comparison on MM-WHS 2017 Dataset.}
\label{fig:results}
\end{figure}

In Figure~\ref{fig:results}, we can observe the whole heart segmentation results of our proposed RL-U$^2$Net method and the latest SOTA segmentation model in both CT and MRI modes of MM-WHS 2017. Compared with other models, our segmentation results are closer to the real ones.

\subsection{Ablation Studies}

To thoroughly validate the effectiveness of each key component of RL-U$^2$Net, we designed a series of ablation experiments to assess the contribution of each core module to the overall segmentation performance by removing them one by one (see Table~\ref{table:ablation} for details). For more detailed ablation experiment contents, please refer to the appendix.

\begin{table}[t]
\centering
\begin{tabular}{ccc}
\hline
\multirow{2}*{Method} & \multicolumn{2}{c}{Dice$\uparrow$} \\
\cline{2-3}
~ & CT & MRI \\
\hline
\textbf{Our model} & \textbf{0.931} & \textbf{0.870} \\
Our model (w/o CMA) & 0.884 & 0.612 \\
Our model (w/o RL-XAlign) & 0.909 & 0.837 \\
Our model (w/o Auxiliary loss) & 0.912 & 0.834 \\
Our model (w/o AGWD) & 0.885 & 0.768 \\
\hline
\end{tabular}
\caption{Ablation studies of diffrent modules and methods in RL-U$^2$Net.}
\label{table:ablation}
\end{table}

\subsubsection{Ablation Experiments for CMA}
Removing the CMA module yields asymmetric performance impacts: The CT segmentation results fluctuate slightly with Dice coefficient declining to 88\%, while MRI performance degrades severely to 60\%. Without semantic correspondence guidance, auxiliary modal features integrate chaotically into the primary modal space, severely disrupting network decisions. This feature confusion particularly affects MRI due to its inherently lower imaging contrast, confirming the critical role of cross-modal attention in structured feature fusion.

\subsubsection{Ablation Experiments for RL-XAlign}
\begin{figure}[t]
\centering
\includegraphics[width=0.99\columnwidth]{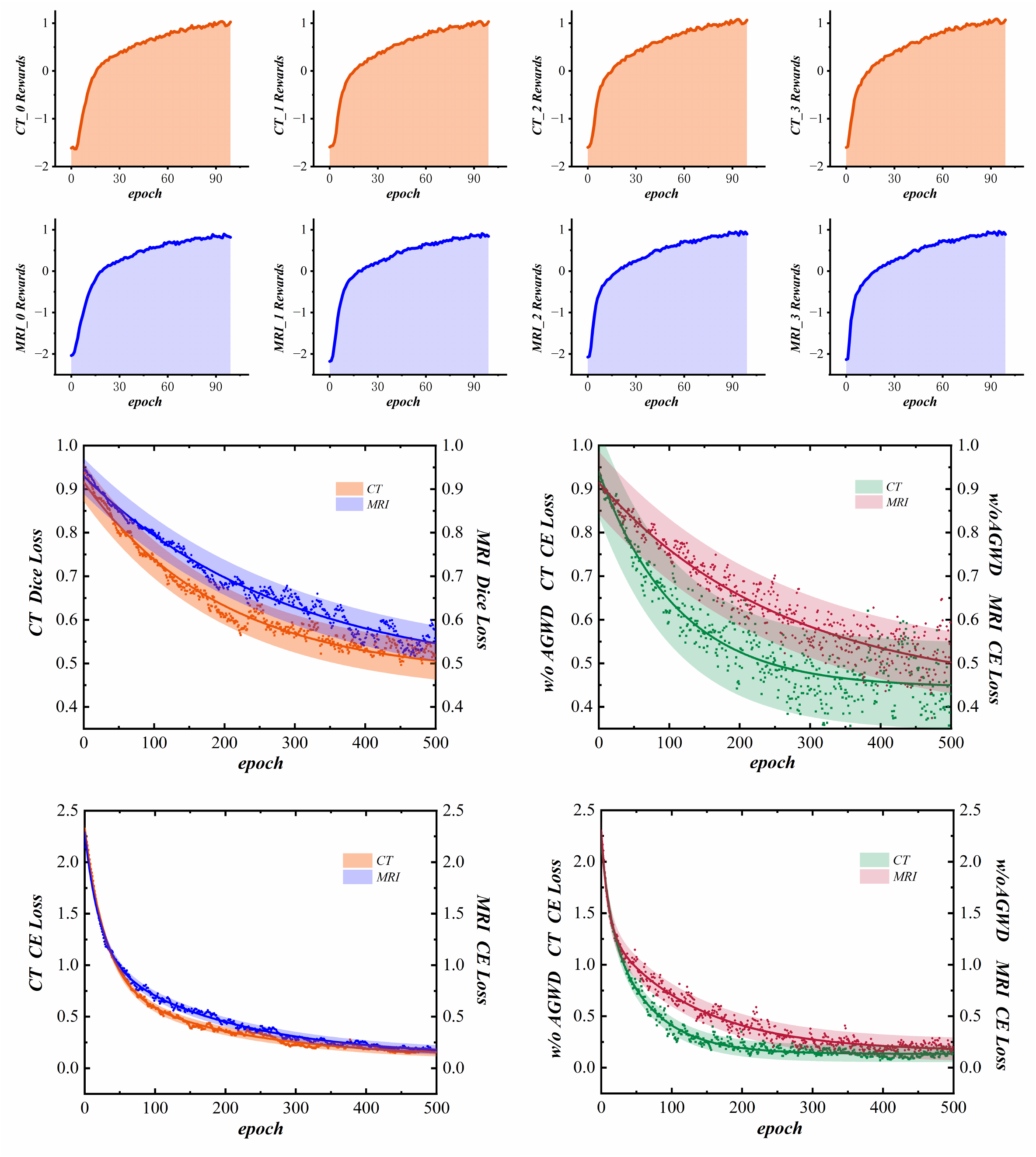}
\caption{Ablation experiments of RL-XAlign and AGWD. The top rows show RL-XAlign reward curves while bottom rows compare loss dynamics with/without AGWD.}
\label{fig:rewards-loss}
\end{figure}

Removing the RL-XAlign module reduces CT and MRI Dice coefficients to 90\% and 83\%, respectively. This degradation stems from the loss of cross-modal interaction, causing the dual-branch network to degenerate into two independent single-modal UNets. Each branch then relies solely on its own modality's limited information, unable to exploit complementary features from the other modality, confirming cross-modal interaction's critical role in segmentation performance. Figure~\ref{fig:rewards-loss} shows PPO reward curves during the first 100 training epochs, where both modalities exhibit steady upward trends, demonstrating successful learning of effective alignment strategies that continuously improve spatial consistency and semantic correspondence.

\subsubsection{Ablation Experiments for Auxiliary Loss}
After removing the auxiliary alignment loss, the performance of CT and MRI decreased slightly to approximately 91\% and 83\%, respectively, with the smallest but still observable decrease. This result indicates that although the auxiliary loss function is not a decisive factor, it plays a significant regularization role in the fine-tuning of feature alignment, effectively constraining the direction of cross-modal feature learning.

\subsubsection{Ablation Experiments for AGWD}
Removing the AGWD reduces CT and MRI performance to 88\% and 76\%, respectively, with MRI showing greater degradation. Figure~\ref{fig:rewards-loss} illustrates that AGWD maintains balanced Dice and cross-entropy losses across modalities with stable convergence. Without AGWD, significant imbalance emerges: CT loss decreases rapidly while MRI converges slowly with large fluctuations, confirming AGWD's effectiveness in addressing multimodal training imbalance.

\section{Conclusion}
We propose RL-U$^2$Net, a dual-branch network leveraging reinforcement learning for multimodal feature fusion in 3D whole-heart segmentation. The RL-XAlign module employs cross-modal attention and RL agents to achieve optimal spatial alignment, addressing multimodal spatial inconsistencies. The AGWD ensures training stability through dynamic modality balancing, while ensemble-based decision fusion enhances prediction accuracy. Comprehensive validation on MM-WHS 2017 achieves 93.15\% and 86.96\% Dice coefficients for CT and MRI, respectively. Experimental results and ablation studies confirm RL-U$^2$Net's superiority over state-of-the-art methods, providing an effective solution for complex medical image analysis.

\appendix
\section{Supplementary Contents}
\subsection{Overview of RL-U$^2$Net structure}
\begin{figure}[t]
\centering
\includegraphics[width=0.99\columnwidth]{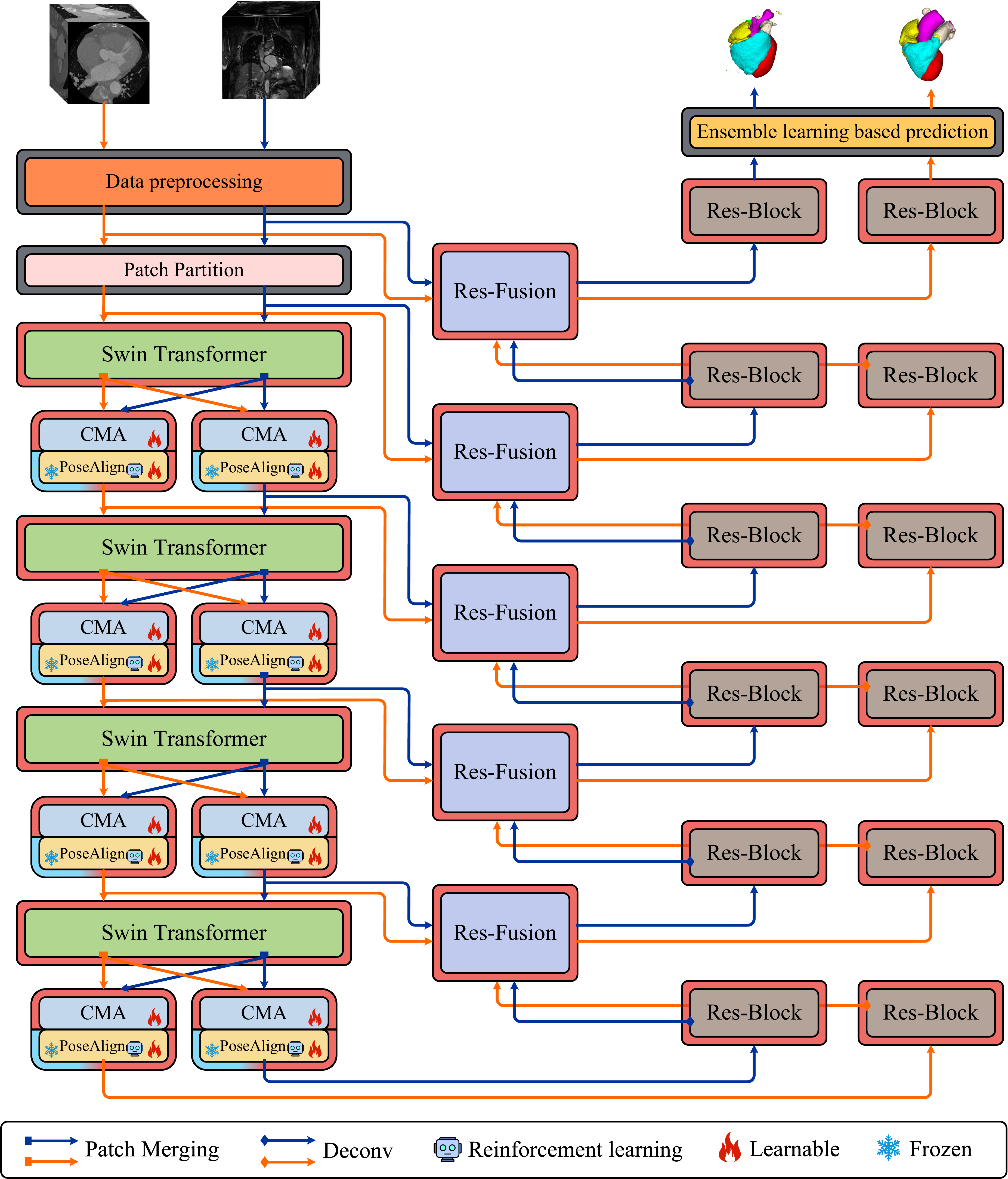}
\caption{Overview of RL-U$^2$Net pipeline.}
\label{fig:RL-U2Net-structure}
\end{figure}

\begin{figure}[t]
\centering
\includegraphics[width=0.99\columnwidth]{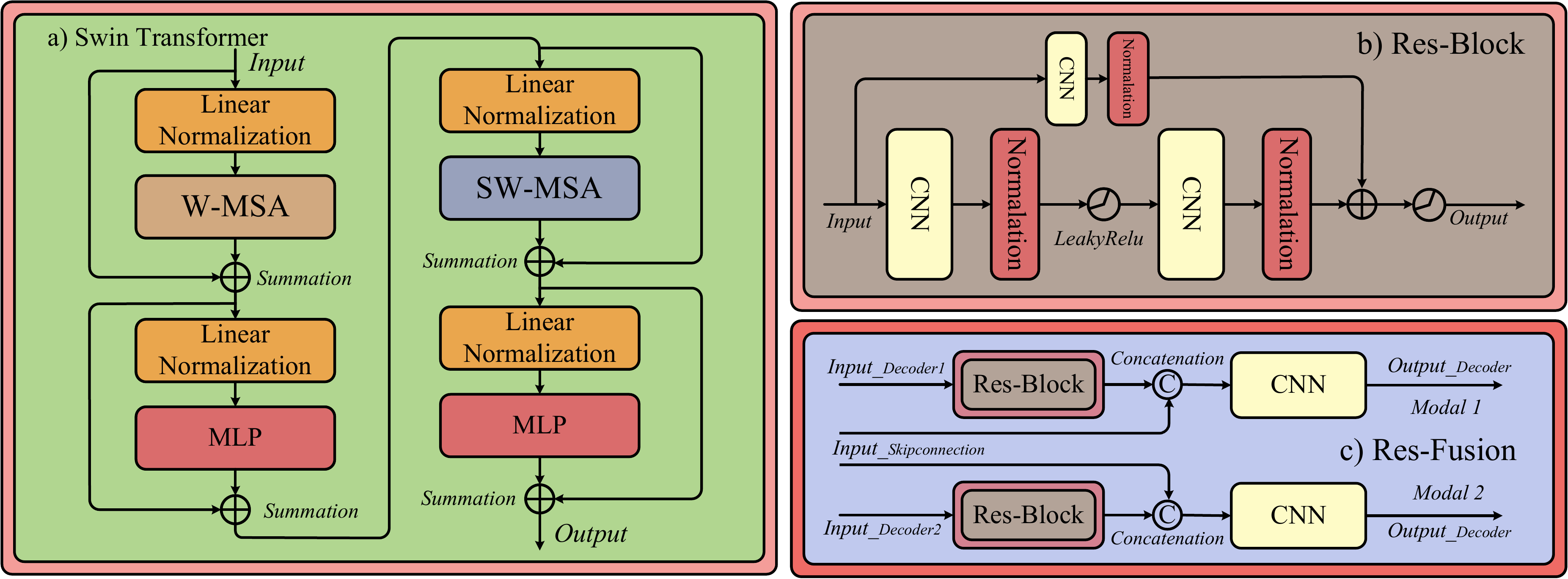}
\caption{Architecture of Swin Transformer, Res-Block and Res-Fusion.}
\label{fig:module-structure}
\end{figure}

Due to visual limitation, in the main text, we only give a brief description of the backbone network. In order to show the structure of the network and related modules more clearly, we provide a more detailed description of the overall architecture of RL-U$^2$Net in the supplementary material. The overall architecture of the proposed RL-U$^2$Net is shown in Figure~\ref{fig:RL-U2Net-structure}. The main network of RL-U$^2$Net consists of a shared encoder based on Swin Transformer, an RL-XAlign cross-modal alignment module, a Res-Fusion fusion module, and a dual-branch ResU-Net decoder. For the input CT and MRI images $I_{\rm{CT}},I_{\rm{MRI}}\in\mathbb{R}^{H\times W\times D}$, the data first undergoes standardization and enhancement through a data preprocessing module, followed by patch segmentation to convert the 3D images into overlapping patch sequences as the network's input. The encoder adopts a hierarchical design consisting of four Swin Transformer stages (Figure~\ref{fig:module-structure}a)), with an RL-XAlign module embedded after each stage to achieve cross-modal feature alignment. The Swin Transformer module is responsible for extracting local-global multi-scale feature representations, while the RL-XAlign module achieves cross-modal feature alignment to address spatial inconsistency issues. At the end of each encoder stage, a Patch Merging operation is performed for downsampling, halving the spatial resolution of the feature map while doubling the channel dimension, thereby progressively constructing a hierarchical feature representation. Inspired by the U-Net architecture, we designed two independent and symmetric ResBlock decoder branches~\cite{he2016deep} (Figure~\ref{fig:module-structure}b)), corresponding to the segmentation tasks for CT and MRI modalities, respectively. Each decoder consists of five upsampling stages, which restore the feature map resolution through deconvolution layers. Each upsampling doubles the spatial size while halving the number of channels. Jump connections are established between the encoder and decoder, and the Res-Fusion module (see Figure~\ref{fig:module-structure}c)) effectively fuses the high-resolution shallow features of the encoder with the high-semantic deep features of the decoder to compensate for the loss of detail information during the downsampling process. Finally, the ensemble learning decision module generates precise whole-heart segmentation results through sliding window inference and weighted voting mechanisms. The following sections will detail the design principles and implementation mechanisms of each core module.

\subsection{Pseudocode of PPO training in RL-XAlign module}

To provide comprehensive implementation details of how reinforcement learning integrates with the backbone network training in RL-XAlign, we have added a description of the training process in the supplementary material.
The training procedure follows a three-phase iterative framework designed to integrate reinforcement learning with standard deep learning optimization. The pseudocode is shown in Algorithm~\ref{alg:ppo-rlu2net}.The experience collection phase systematically gathers training samples by evaluating both baseline and policy-guided states for each CT-MRI pair.  During this phase, the algorithm first disables the PoseAlign module to establish baseline measurements of Dice coefficient and similarity scores, then re-enables the module to capture policy-driven outcomes and corresponding log probabilities.  Each collected experience tuple encompasses current state vectors, next state vectors, policy actions, and computed rewards based on segmentation performance improvements.  The policy optimization phase implements PPO's clipped surrogate objective through multiple mini-epochs, where shuffled experience batches enable stable policy updates via advantage estimation and importance ratio clipping.  This phase alternates between policy network updates using the clipped loss and value network refinements through mean squared error minimization.  The backbone optimization phase concludes each training iteration by updating the underlying segmentation network parameters through conventional gradient descent, ensuring that feature extraction capabilities evolve alongside the alignment policies.  This unified framework enables end-to-end learning where reinforcement learning-guided feature alignment and deep learning-based segmentation mutually enhance each other throughout the training process.

\begin{algorithm}[tbhp]
\caption{PPO Training with RL-U$^2$Net on Multimodal Dataset}
\label{alg:ppo-rlu2net}
\textbf{Input}: Multimodal dataset $\mathcal{D}=\{(\texttt{ct},\texttt{mr})\}$, actor \texttt{model} (RL-U$^2$Net), critic \texttt{critic}\\
\textbf{Parameter}: samples $K$, discount $\gamma$, clip range $\epsilon$, temperature $\tau$, PPO mini-epochs $M$, base deltas $\Delta_{\text{Dice}}$, $\Delta_{\text{CE}}$\\
\textbf{Output}: NA
\begin{algorithmic}[1] 
\STATE initialize PPO buffer $\mathcal{B}\gets\emptyset$.

\STATE \textit{// Data collection (rollouts)}
\FORALL{$(\texttt{ct},\texttt{mr}) \in \mathcal{D}$}
\FOR{$k \gets 1$ \TO $K$}
  \STATE $\phi_{\text{now}} \gets e/E$
  \STATE Disable PoseAlign.
  \STATE $(d_{\text{now}}, s_{\text{now}}) \gets \texttt{model.forward}(\texttt{ct}, \texttt{mr})$
  \STATE Restore PoseAlign.
  \STATE $x_{\text{now}} \gets (d_{\text{now}}, s_{\text{now}}, \phi_{\text{now}})$
  \STATE $\phi_{\text{next}} \gets (e+1)/E$
  \STATE $(d_{\text{next}}, s_{\text{next}}, \log p_{\text{old}}, \pi)\gets $\\
  $\texttt{model.forward}(\texttt{ct}, \texttt{mr}; \tau)$
  \STATE $x_{\text{next}} \gets (d_{\text{next}}, s_{\text{next}}, \phi_{\text{next}})$
  \STATE $r \gets d_{\text{next}} + s_{\text{next}}$
  \STATE $\mathcal{B}\gets \mathcal{B}\cup\{(x_{\text{now}}, x_{\text{next}}, \log p_{\text{old}}, \pi, r)\}$
\ENDFOR
\ENDFOR

\STATE \textit{// PPO update}
\FOR{$m \gets 1$ \TO $M$}
\FORALL{$(x_{\text{now}}, x_{\text{next}}, \log p_{\text{old}}, \pi, r)\in \mathcal{B}$}
  \STATE $V_{\text{now}} \gets \texttt{critic.forward}(x_{\text{now}})$
  \STATE $V_{\text{next}} \gets \texttt{critic.forward}(x_{\text{next}})$
  \STATE $\hat V \gets r + \gamma\cdot V_{\text{next}}$
  \STATE $A \gets \hat V - V_{\text{now}}$
  \STATE $\log p_{\text{new}} \gets \texttt{model.get\_log\_prob}(\pi)$
  \STATE $\rho \gets \exp(\log p_{\text{new}}-\log p_{\text{old}})$
  \STATE $L^{\text{CLIP}} \gets \min\big(\rho A,\ \texttt{clip}(\rho,1-\epsilon,1+\epsilon)\cdot A\big)$
  \STATE $\mathcal{L}_{\text{actor}} \gets -\texttt{mean}(L^{\text{CLIP}})$
  \STATE $\mathcal{L}_{\text{critic}} \gets \texttt{MSE}(V_{\text{now}}, \hat V)$
  \STATE \texttt{model.update.reinforcelearning}
  \STATE \texttt{critic.update}
\ENDFOR
\ENDFOR

\STATE \textit{// Supervised deep learning head}
\FORALL{$(\texttt{ct},\texttt{mr}) \in \mathcal{D}$}
\STATE \texttt{model.update.deeplearning}()
\ENDFOR

\STATE $\mathcal{B}\gets\emptyset$;\quad $e\gets e+1$

\end{algorithmic}
\end{algorithm}

\begin{table*}[t]
\centering
\begin{tabular}{cccccccccc}
\hline
\multirow{2}*{Method} & \multicolumn{8}{c}{Dice$\uparrow$} & \multirow{2}*{HD95(mm)$\downarrow$} \\
\cline{2-9}
~ & Myo & LA & LV & RA & RV & AA & PA & Average & ~ \\
\hline
3D U-Net~\cite{cicek20163d} & 0.686 & 0.844 & 0.873 & 0.824 & 0.808 & 0.783 & 0.755 & 0.796 & 39.996 \\
ConResNet~\cite{lee20223d} & 0.848 & 0.887 & 0.922 & 0.865 & 0.847 & 0.811 & 0.775 & 0.851 & 42.411 \\
nnformer~\cite{zhou2023nnformer} & 0.682 & 0.814 & 0.848 & 0.846 & 0.824 & 0.787 & 0.763 & 0.795 & 31.557 \\
D-Former~\cite{wu2023dformer} & 0.732 & 0.849 & 0.887 & 0.891 & 0.849 & 0.811 & 0.784 & 0.829 & 31.097 \\
SwinUNETR & \multirow{2}*{0.741} & \multirow{2}*{0.810} & \multirow{2}*{0.880} & \multirow{2}*{0.853} & \multirow{2}*{0.829} & \multirow{2}*{0.792} & \multirow{2}*{0.792} & \multirow{2}*{0.814} & \multirow{2}*{40.017} \\
\cite{hatamizadeh2021swin}\\
UNETR++~\cite{shaker2024unetr} & 0.711 & 0.829 & 0.883 & 0.883 & 0.848 & 0.807 & 0.813 & 0.825 & 30.240 \\
Cascaded U-Net & \multirow{2}*{0.810} & \multirow{2}*{0.889} & \multirow{2}*{0.936} & \multirow{2}*{0.884} & \multirow{2}*{\textbf{0.898}} & \multirow{2}*{0.819} & \multirow{2}*{0.815} & \multirow{2}*{0.864} & \multirow{2}*{29.629} \\
\cite{salgado-garcia2024efficient}\\
HRMedSeg~\cite{xu2025hrmedseg} & 0.843 & 0.885 & 0.933 & 0.894 & 0.877 & 0.815 & \textbf{0.827} & 0.868 & \textbf{27.616} \\
P2TC~\cite{cui2025p2tc} & 0.837 & 0.890 & 0.928 & \textbf{0.896} & 0.892 & 0.835 & 0.817 & \textbf{0.871} & 45.493 \\
\hline
RL-U$^2$Net(Ours) & \textbf{0.865} & \textbf{0.897} & \textbf{0.945} & 0.863 & 0.859 & \textbf{0.837} & 0.823 & 0.870 & 29.741 \\
\hline
\end{tabular}
\caption{Performance comparison of RL-U$^2$Net and the SOTA segmentation methods on the MM-WHS 2017 MRI dataset.}
\label{table:mr}
\end{table*}

\subsection{Supplements to Datasets and Pre-Processings}
A more detailed description of the dataset and data preprocessing methods are added in this section. In this study, we mainly used the MM-WHS dataset to evaluate the performance of the model we proposed, and then conducted supplementary experiments using the MyoPS 2020 data to verify the generalization performance of the model.

The MM-WHS 2017 dataset used in this study was obtained from the MM-WHS Challenge 2017~\cite{zhuang2019evaluation} [49], which includes 60 sets of cardiac CT and MRI image data. Among these, 20 sets were manually annotated and used as training data, while the remaining 40 sets of unannotated data were used as test data. Since only 20 sets of publicly annotated data were available, the dataset needed to be re-divided to avoid overfitting. Following the data partitioning strategy adopted in previous studies~\cite{cui2023improved,cui2025p2tc}, the 40 sets of data were used as the training set, while the 20 sets of labeled data were randomly divided into 15 sets for testing and 5 sets for validation. To ensure the consistency of the training data and the stable convergence of the model, all images underwent a standardized preprocessing workflow. This includes: resampling images to a uniform voxel spacing of $(1.0, 1.0, 1.5)$ mm to balance resampling quality and label integrity; Performing intensity normalization, where CT image intensity ranges are standardized to $[-175, 250]$ and mapped to $[0, 1]$, and MRI image intensity ranges are standardized to $[150, 1488]$ and mapped to $[0, 1]$.

The MyoPS dataset had 45 cases of multi-sequence CMR (25 cases for training and 20 cases for testing), each of which refers to a patient with three sequence CMR, i.e., LGE, T2 and bSSFP CMR. The data have been pre-processed using the MvMM method~\cite{zhuang2016multivariate,zhuang2018multivariate}, to align the three-sequence CMR into a common space and to resample them into the same spatial resolution. The provided gold-standard labels include LV blood pool, RV blood pool, LV normal myocardium, LV myocardial edema, and LV myocardial scars. To focus on the segmentation of cardiac structure, we combined myocardium, myocardial edema and myocardial scar into a single myocardial category~\cite{yan2021novel}.

\subsection{Training Strategy and Implementation Details}
All experiments in this study were implemented using the PyTorch deep learning framework, with a Python 3.10 environment, and trained on a server equipped with an NVIDIA A10 GPU. The network uses the AdamW optimizer, with a learning rate of $1\times 10^{-4}$ for the backbone network, a weight decay coefficient of $1\times 10^{-5}$, and a momentum parameter of 0.99. Given the complex multi-stage training characteristics of RL-U$^2$Net, a hierarchical training strategy was adopted. The overall training was set to 500 epochs: the first 100 epochs simultaneously conducted reinforcement learning training and backbone network training, with the pose alignment strategy in the RL-XAlign module co-optimized with the main segmentation task; The RL-XAlign module uses a dedicated hyperparameter configuration: the policy network learning rate is set to $5\times 10^{-5}$, and the value network learning rate is set to $2\times 10^{-4}$. Each reinforcement learning round includes 16 environment samples, followed by 10 mini-rounds of policy updates, with a batch size of 64 for each mini-round. The subsequent 400 rounds are dedicated to fine-tuning the backbone network, ensuring that the encoder and decoder converge further on the basis of the optimized feature alignment. The learning rate scheduling adopts a cosine annealing strategy with preheating, with the preheating phase set to 50 rounds, the learning rate linearly increasing from 0 to the set value, and then decaying according to a cosine function until the end of training.

\begin{figure*}[t]
\centering
\includegraphics[width=0.99\textwidth]{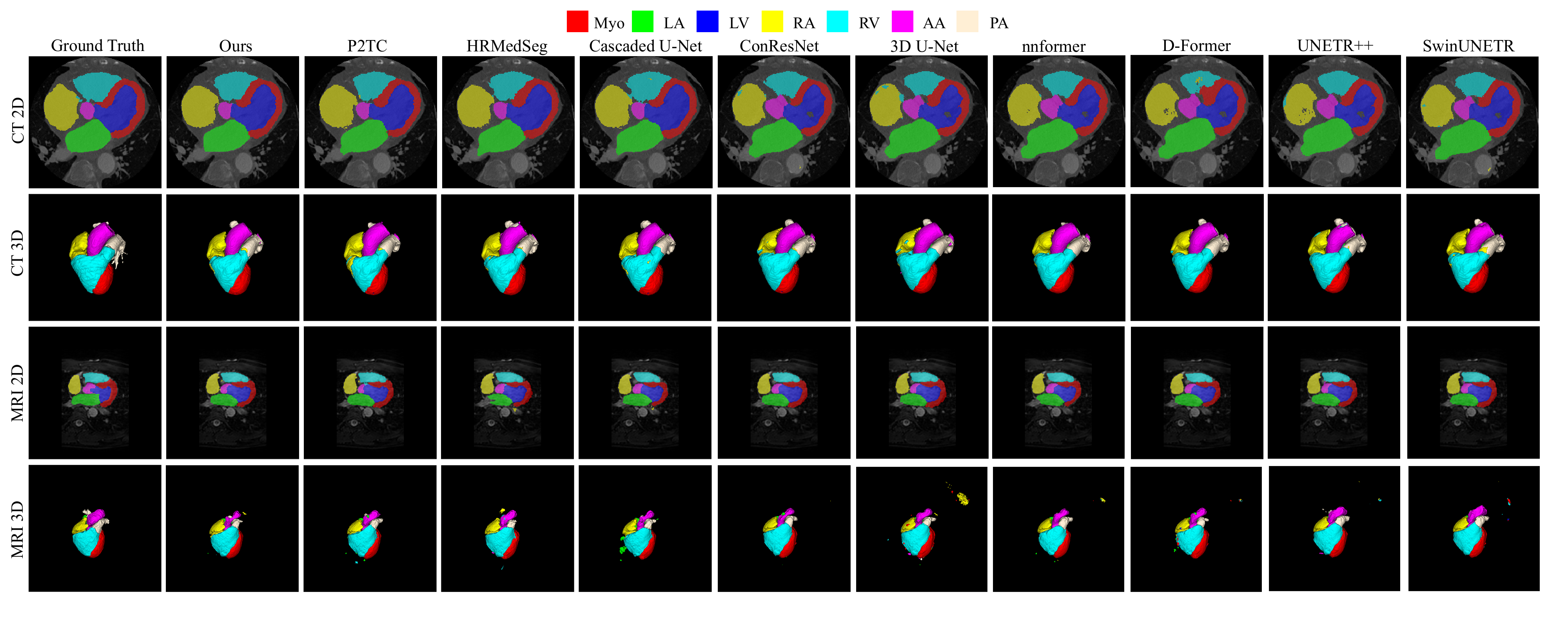}
\caption{Visualization of methods comparison on MM-WHS 2017 Dataset.}
\label{fig:results-all}
\end{figure*}

\subsection{MRI Results on MM-WHS 2017 Challenge Dataset}

Due to visual limitations, we only presented the segmentation results of CT modalities on MM-WHS 2017 in the main text. To comprehensively evaluate the segmentation performance of RL-U$^2$Net, we supplemented the segmentation results of RL-U$^2$Net and 11 state-of-the-art segmentation models on the MM-WHS 2017 MRI dataset, as shown in the Table~\ref{table:mr}.

In MRI image segmentation tasks, RL-U$^2$Net also maintains a leading advantage, with an average Dice coefficient of 87.0\%, although this is 0.1 percentage points lower than P2TC's 87.1\%, it performs more stably on most individual anatomical structures. In terms of Hausdorff distance, RL-U$^2$Net achieves an outstanding performance of 29.741 mm, showing significant improvement compared to most comparison methods.

We have also supplemented the complete comparison figure of the visualization results. In Figure~\ref{fig:results-all}, the visualization results fully demonstrate the segmentation results of different models in CT and MRI modes. RL-U$^2$Net shows consistent advantages in both imaging environments. In CT modality (rows 1-2), our method produces segmentation masks with sharp anatomical boundaries and accurate structural delineation, particularly excelling in complex regions such as the left atrium (LA) where competing methods exhibit noticeable over-segmentation or boundary artifacts. Some comparison methods show varying degrees of structural inconsistencies, with producing fragmented regions or missing fine anatomical details. In MRI modality (rows 3-4), the segmentation task becomes significantly more challenging due to inherent soft tissue contrast limitations, yet RL-U$^2$Net maintains robust performance with well-preserved anatomical topology. Notably, while most comparison methods struggle with MRI's lower signal-to-noise ratio, resulting in irregular boundaries and incomplete structure identification, our approach consistently delivers smooth, anatomically plausible segmentation masks that closely approximate the ground truth annotations. This cross-modal consistency underscores the effectiveness of our reinforcement learning-assisted feature alignment strategy in handling the distinct imaging characteristics and spatial variations inherent to each modality.

\subsection{Results on MyoPS Dataset}
We conducted supplementary experiments on the public MyoPS2020 dataset. This dataset includes three cardiac magnetic resonance (CMR) modalities: bSSFP, LGE, and T2-SPAIR. We implemented three cross-modal configurations—bSSFP+LGE, T2+LGE, and T2+bSSFP—to comprehensively assess the robustness of cross-sequence fusion. The results are shown in Table~\ref{table:MyoPS}.

When bSSFP served as the primary modality, using LGE as auxiliary input yielded an average Dice of 85.73\%  and an HD95 of 6.92. This slightly outperformed the configuration using the T2 modality as auxiliary. When LGE was the primary modality, using bSSFP as auxiliary input achieved the best overall performance in this study, with an average Dice of 87.90\% and an HD95 of 7.95. Finally, when T2-SPAIR was used as the main source and bSSFP was used as the auxiliary source to obtain Avg Dice=86.59\% and HD95=8.18; The results of LGE as an auxiliary source were Avg Dice= 86.01\% and HD95=9.53.

These supplementary results demonstrate that our proposed RL-U2Net framework achieves consistently robust and effective segmentation performance across diverse cross-modal combinations on the MyoPS2020 dataset. The bSSFP modality, in particular, demonstrated significant informational value, contributing to strong performance both as a primary and an auxiliary source.

\begin{table}[t]
\centering

\begin{tabular}{cccccc}
\hline
Auxiliary & \multicolumn{4}{c}{Dice $\uparrow$} & \multirow{2}*{HD95 $\downarrow$} \\ 

modality & Myo & LV & RV & Avg & ~ \\
\hline

\multicolumn{6}{c}{bSSFP cine CMR results} \\
\hline
LGE & 83.06 & 85.49 & 88.63 & 85.73 & 6.92 \\
T2 & 82.52 & 85.23 & 87.78 & 85.18 & 7.59 \\
\hline

\multicolumn{6}{c}{LGE CMR results} \\
\hline
bSSFP & 87.22 & 88.17 & 88.30 & 87.90 & 7.95 \\
T2 & 85.84 & 87.95 & 87.37 & 87.05 & 11.49 \\
\hline

\multicolumn{6}{c}{T2-SPAIR CMR results} \\
\hline
bSSFP & 87.48 & 87.14 & 85.14 & 86.59 & 8.18 \\
LGE & 86.80 & 86.51 & 84.74 & 86.01 & 9.53 \\
\hline
\end{tabular}
\caption{Supplementary experiments on the MyoPS dataset.}
\label{table:MyoPS}
\end{table}

\subsection{Supplementary Ablation Experiment}
In the main text, we performed ablation experiments to investigate the effects of the core components of the model, the CMA module, the RL-XAlign module, the Auxiliary loss and the AGWD, on the model segmentation performance. In this Supplementary Material, we further evaluate the impact of the Ensemble-based Decision Module and Fusion weight on the model performance.

\subsubsection{Ablation Experiments for Ensemble-based Decision Module}
Removing the ensemble-based prediction module results in moderate performance degradation, with CT and MRI Dice coefficients declining to 91.9\% and 86.1\%, respectively. Correspondingly, Hausdorff distances increase to 15.493mm for CT and 42.515mm for MRI, indicating compromised boundary precision. While this module contributes modestly to overall accuracy, it effectively addresses the critical challenge of aggregating patch-level predictions into coherent whole-image segmentation, particularly important for maintaining spatial consistency across overlapping regions in multimodal fusion scenarios.

\subsubsection{Fusion Weight Analysis for Cross-modal Integration}

\begin{figure}[t]
\centering
\includegraphics[width=0.99\columnwidth]{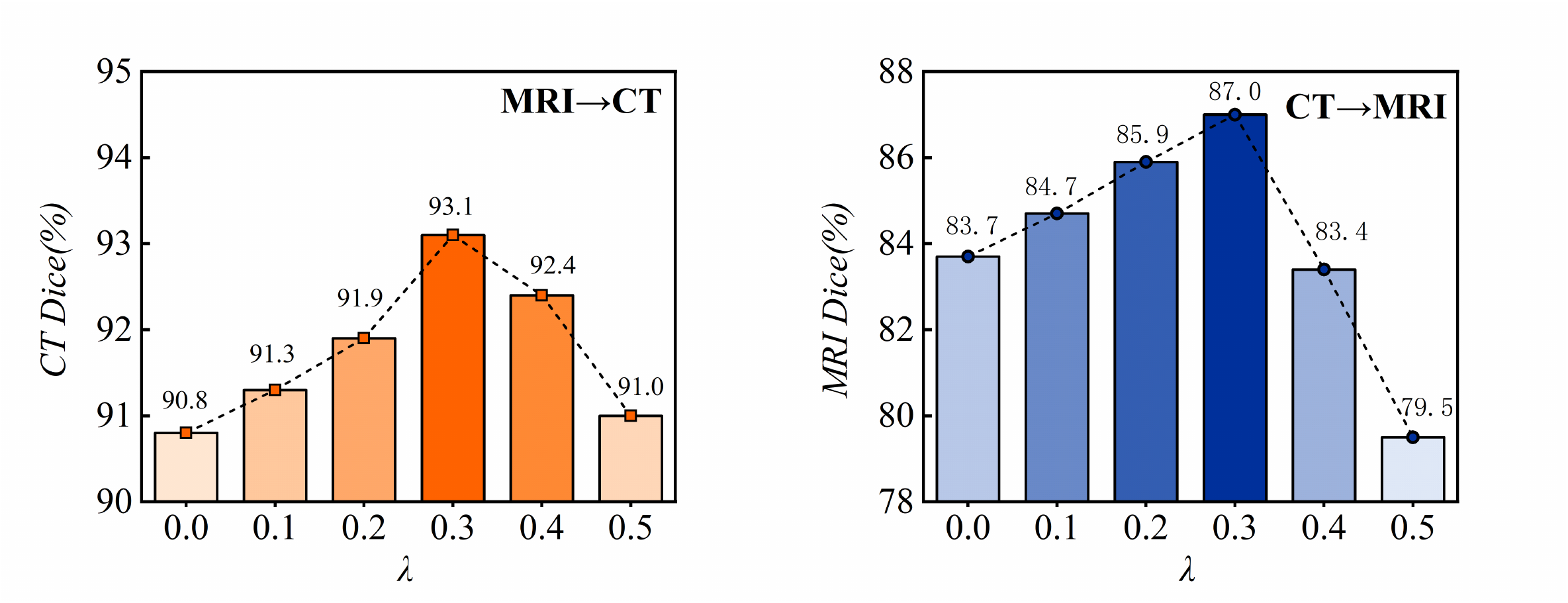}
\caption{Analysis of the fusion weight $\lambda$. The left figure shows the segmentation results under different $\lambda$ with CT as the main mode and MRI as the auxiliary mode. The right figure shows the segmentation results under different $\lambda$ when MRI is used as the main mode and CT is used as the auxiliary mode}
\label{fig:analysis-lambda}
\end{figure}

We systematically investigate the impact of fusion weight $\lambda$(Equation(7) of main text)in the cross-modal feature integration process, the dice results of different $\lambda$ are shown in the Figure~\ref{fig:analysis-lambda}. When CT serves as the primary modality with MRI auxiliary features, performance peaks at $\lambda=0.3$ with 93.1\% Dice coefficient, declining notably at both extremes. Similarly, with MRI as primary and CT auxiliary, optimal performance occurs at $\lambda=0.3$ yielding 87.0\%, while higher fusion weights cause significant degradation. This consistent optimal point at $\lambda=0.3$ across both modality configurations demonstrates that moderate auxiliary feature integration preserves primary modality dominance while effectively leveraging complementary cross-modal information, validating our architectural design choice for balanced multimodal fusion.

\bibliography{reference}

@inproceedings{alansary2018automatic,
  author = {Alansary, Amir and Folgoc, Loic Le and Vaillant, Ghislain and Oktay, Ozan and Li, Yuanwei and {et al.}},
  title = {Automatic View Planning with Multi-Scale Deep Reinforcement Learning Agents},
  pages = {277--285},
  year = {2018},
  booktitle = {International Conference on Medical Image Computing and Computer-Assisted Intervention},
  publisher = {Springer}
}

@article{arulkumaran2017brief,
  author = {Arulkumaran, Kai and Deisenroth, Marc Peter and Brundage, Miles and Bharath, Anil Anthony},
  title = {A Brief Survey of Deep Reinforcement Learning},
  journal = {arXiv preprint arXiv:1708.05866},
  year = {2017},
  eprint = {1708.05866},
  archiveprefix = {arXiv}
}

@inproceedings{cao2022swinunet,
  author = {Cao, Hu and Wang, Yueyue and Chen, Joy and Jiang, Dongsheng and Zhang, Xiaopeng and Tian, Qi and Wang, Manning},
  title = {Swin-Unet: Unet-like Pure Transformer for Medical Image Segmentation},
  pages = {205--218},
  year = {2022},
  booktitle = {European Conference on Computer Vision},
  publisher = {Springer}
}

@inproceedings{carion2020endtoend,
  author = {Carion, Nicolas and Massa, Francisco and Synnaeve, Gabriel and Usunier, Nicolas and Kirillov, Alexander and Zagoruyko, Sergey},
  title = {End-to-End Object Detection with Transformers},
  pages = {213--229},
  year = {2020},
  booktitle = {European Conference on Computer Vision},
  publisher = {Springer}
}

@inproceedings{chang2024emnet,
  author = {Chang, Ao and Zeng, Jiajun and Huang, Ruobing and Ni, Dong},
  title = {Em-Net: Efficient Channel and Frequency Learning with Mamba for 3d Medical Image Segmentation},
  pages = {266--275},
  year = {2024},
  booktitle = {International Conference on Medical Image Computing and Computer-Assisted Intervention},
  publisher = {Springer}
}

@article{chen2021transunet,
  author = {Chen, Jieneng and Lu, Yongyi and Yu, Qihang and Luo, Xiangde and Adeli, Ehsan and {et al.}},
  title = {Transunet: Transformers Make Strong Encoders for Medical Image Segmentation},
  journal = {arXiv preprint arXiv:2102.04306},
  year = {2021},
  eprint = {2102.04306},
  archiveprefix = {arXiv}
}

@inproceedings{cicek20163d,
  author = {{\c C}i{\c c}ek, {\"O}zg{\"u}n and Abdulkadir, Ahmed and Lienkamp, Soeren S. and Brox, Thomas and Ronneberger, Olaf},
  title = {3D U-Net: Learning Dense Volumetric Segmentation from Sparse Annotation},
  pages = {424--432},
  year = {2016},
  booktitle = {International Conference on Medical Image Computing and Computer-Assisted Intervention},
  publisher = {Springer}
}

@article{cui2023improved,
  author = {Cui, Hengfei and Wang, Yifan and Li, Yan and Xu, Di and Jiang, Lei and Xia, Yong and Zhang, Yanning},
  title = {An Improved Combination of Faster R-CNN and U-Net Network for Accurate Multi-Modality Whole Heart Segmentation},
  journal = {IEEE journal of biomedical and health informatics},
  volume = {27},
  number = {7},
  pages = {3408--3419},
  year = {2023},
  publisher = {IEEE},
  isbn = {2168-2194}
}

@article{cui2025p2tc,
  author = {Cui, Hengfei and Wang, Yifan and Zheng, Fan and Li, Yan and Zhang, Yanning and Xia, Yong},
  title = {P2TC: A Lightweight Pyramid Pooling Transformer-CNN Network for Accurate 3D Whole Heart Segmentation},
  journal = {IEEE Journal of Biomedical and Health Informatics},
  year = {2025},
  publisher = {IEEE},
  isbn = {2168-2194}
}

@article{freed2016mr,
  author = {Freed, Benjamin H. and Collins, Jeremy D. and Fran{\c c}ois, Christopher J. and Barker, Alex J. and Cuttica, Michael J. and {et al.}},
  title = {MR and CT Imaging for the Evaluation of Pulmonary Hypertension},
  journal = {JACC: Cardiovascular Imaging},
  volume = {9},
  number = {6},
  pages = {715--732},
  year = {2016},
  publisher = {American College of Cardiology Foundation Washington, DC},
  isbn = {1936-878X}
}

@inproceedings{ghesu2016artificial,
  author = {Ghesu, Florin C. and Georgescu, Bogdan and Mansi, Tommaso and Neumann, Dominik and Hornegger, Joachim and Comaniciu, Dorin},
  title = {An Artificial Agent for Anatomical Landmark Detection in Medical Images},
  pages = {229--237},
  year = {2016},
  booktitle = {International Conference on Medical Image Computing and Computer-Assisted Intervention},
  publisher = {Springer}
}

@inproceedings{hatamizadeh2021swin,
  author = {Hatamizadeh, Ali and Nath, Vishwesh and Tang, Yucheng and Yang, Dong and Roth, Holger R. and Xu, Daguang},
  title = {Swin Unetr: Swin Transformers for Semantic Segmentation of Brain Tumors in Mri Images},
  pages = {272--284},
  year = {2021},
  booktitle = {International MICCAI Brainlesion Workshop},
  publisher = {Springer}
}

@inproceedings{hatamizadeh2022unetr,
  author = {Hatamizadeh, Ali and Tang, Yucheng and Nath, Vishwesh and Yang, Dong and Myronenko, Andriy and {et al.}},
  title = {Unetr: Transformers for 3d Medical Image Segmentation},
  pages = {574--584},
  year = {2022},
  booktitle = {Proceedings of the IEEE/CVF Winter Conference on Applications of Computer Vision}
}

@inproceedings{he2016deep,
  author = {He, Kaiming and Zhang, Xiangyu and Ren, Shaoqing and Sun, Jian},
  title = {Deep Residual Learning for Image Recognition},
  pages = {770--778},
  year = {2016},
  booktitle = {Proceedings of the IEEE Conference on Computer Vision and Pattern Recognition}
}

@article{huang2025highresolution,
  author = {Huang, Xinyu and Dong, Yuhao and Tian, Weiwei and Li, Bo and Feng, Rui and Liu, Ziwei},
  title = {High-Resolution Visual Reasoning via Multi-Turn Grounding-Based Reinforcement Learning},
  journal = {arXiv preprint arXiv:2507.05920},
  year = {2025},
  eprint = {2507.05920},
  archiveprefix = {arXiv}
}

@article{jin2020raunet,
  author = {Jin, Qiangguo and Meng, Zhaopeng and Sun, Changming and Cui, Hui and Su, Ran},
  title = {RA-UNet: A Hybrid Deep Attention-Aware Network to Extract Liver and Tumor in CT Scans},
  journal = {Frontiers in Bioengineering and Biotechnology},
  volume = {8},
  pages = {605132},
  year = {2020},
  publisher = {Frontiers Media SA},
  isbn = {2296-4185}
}

@article{kaelbling1996reinforcement,
  author = {Kaelbling, Leslie Pack and Littman, Michael L. and Moore, Andrew W.},
  title = {Reinforcement Learning: A Survey},
  journal = {Journal of artificial intelligence research},
  volume = {4},
  pages = {237--285},
  year = {1996},
  isbn = {1076-9757}
}

@article{koehler2021unsupervised,
  author = {Koehler, Sven and Hussain, Tarique and Blair, Zach and Huffaker, Tyler and Ritzmann, Florian and {et al.}},
  title = {Unsupervised Domain Adaptation from Axial to Short-Axis Multi-Slice Cardiac MR Images by Incorporating Pretrained Task Networks},
  journal = {IEEE transactions on medical imaging},
  volume = {40},
  number = {10},
  pages = {2939--2953},
  year = {2021},
  publisher = {IEEE},
  isbn = {0278-0062}
}

@article{lee20223d,
  author = {Lee, Ho Hin and Bao, Shunxing and Huo, Yuankai and Landman, Bennett A.},
  title = {3d Ux-Net: A Large Kernel Volumetric Convnet Modernizing Hierarchical Transformer for Medical Image Segmentation},
  journal = {arXiv preprint arXiv:2209.15076},
  year = {2022},
  eprint = {2209.15076},
  archiveprefix = {arXiv}
}

@article{li2022myocardial,
  author = {Li, Weisheng and Wang, Linhong and Li, Feiyan and Qin, Sheng and Xiao, Bin},
  title = {Myocardial Pathology Segmentation of Multi-Modal Cardiac MR Images with a Simple but Efficient Siamese U-Shaped Network},
  journal = {Biomedical Signal Processing and Control},
  volume = {71},
  pages = {103174},
  year = {2022},
  publisher = {Elsevier},
  isbn = {1746-8094}
}

@article{li2023multimodality,
  author = {Li, Lei and Ding, Wangbin and Huang, Liqin and Zhuang, Xiahai and Grau, Vicente},
  title = {Multi-Modality Cardiac Image Computing: A Survey},
  journal = {Medical image analysis},
  volume = {88},
  pages = {102869},
  year = {2023},
  publisher = {Elsevier},
  isbn = {1361-8415}
}

@article{li2023myops,
  author = {Li, Lei and Wu, Fuping and Wang, Sihan and Luo, Xinzhe and {Mart{\'i}n-Isla}, Carlos and {et al.}},
  title = {MyoPS: A Benchmark of Myocardial Pathology Segmentation Combining Three-Sequence Cardiac Magnetic Resonance Images},
  journal = {Medical Image Analysis},
  volume = {87},
  pages = {102808},
  year = {2023},
  publisher = {Elsevier},
  isbn = {1361-8415}
}

@inproceedings{liao2020iterativelyrefined,
  author = {Liao, Xuan and Li, Wenhao and Xu, Qisen and Wang, Xiangfeng and Jin, Bo and {et al.}},
  title = {Iteratively-Refined Interactive 3D Medical Image Segmentation with Multi-Agent Reinforcement Learning},
  pages = {9394--9402},
  year = {2020},
  booktitle = {Proceedings of the IEEE/CVF Conference on Computer Vision and Pattern Recognition}
}

@article{liu2025pixel,
  author = {Liu, Yunxin and Yuan, Di and Xu, Zhenghua and Zhan, Yuefu and Zhang, Hongwei and Lu, Jun and Lukasiewicz, Thomas},
  title = {Pixel Level Deep Reinforcement Learning for Accurate and Robust Medical Image Segmentation},
  journal = {Scientific Reports},
  volume = {15},
  number = {1},
  pages = {8213},
  year = {2025},
  publisher = {Nature Publishing Group UK London},
  isbn = {2045-2322}
}

@inproceedings{luo2020mvmmregnet,
  author = {Luo, Xinzhe and Zhuang, Xiahai},
  title = {MvMM-RegNet: A New Image Registration Framework Based on Multivariate Mixture Model and Neural Network Estimation},
  pages = {149--159},
  year = {2020},
  booktitle = {International Conference on Medical Image Computing and Computer-Assisted Intervention},
  publisher = {Springer}
}

@article{luo2022mathcal,
  author = {Luo, Xinzhe and Zhuang, Xiahai},
  title = {X-Metric: An N-Dimensional Information-Theoretic Framework for Groupwise Registration and Deep Combined Computing},
  journal = {IEEE Transactions on Pattern Analysis and Machine Intelligence},
  volume = {45},
  number = {7},
  pages = {9206--9224},
  year = {2022},
  publisher = {IEEE},
  isbn = {0162-8828}
}

@article{oord2018representation,
  author = {van den Oord, Aaron and Li, Yazhe and Vinyals, Oriol},
  title = {Representation Learning with Contrastive Predictive Coding},
  journal = {arXiv preprint arXiv:1807.03748},
  year = {2018},
  eprint = {1807.03748},
  archiveprefix = {arXiv}
}

@article{pan2025multiscale,
  author = {Pan, Peng and Zhang, Chengxue and Sun, Jingbo and Guo, Lina},
  title = {Multi-Scale Conv-Attention U-Net for Medical Image Segmentation},
  journal = {Scientific Reports},
  volume = {15},
  number = {1},
  pages = {12041},
  year = {2025},
  publisher = {Nature Publishing Group UK London},
  isbn = {2045-2322}
}

@article{payer2019segmenting,
  author = {Payer, Christian and {\v S}tern, Darko and Feiner, Marlies and Bischof, Horst and Urschler, Martin},
  title = {Segmenting and Tracking Cell Instances with Cosine Embeddings and Recurrent Hourglass Networks},
  journal = {Medical image analysis},
  volume = {57},
  pages = {106--119},
  year = {2019},
  publisher = {Elsevier},
  isbn = {1361-8415}
}

@article{pei2021disentangle,
  author = {Pei, Chenhao and Wu, Fuping and Huang, Liqin and Zhuang, Xiahai},
  title = {Disentangle Domain Features for Cross-Modality Cardiac Image Segmentation},
  journal = {Medical Image Analysis},
  volume = {71},
  pages = {102078},
  year = {2021},
  publisher = {Elsevier},
  isbn = {1361-8415}
}

@article{puyol-anton2022multimodal,
  author = {{Puyol-Ant{\'o}n}, Esther and Sidhu, Baldeep S. and Gould, Justin and Porter, Bradley and Elliott, Mark K. and {et al.}},
  title = {A Multimodal Deep Learning Model for Cardiac Resynchronisation Therapy Response Prediction},
  journal = {Medical Image Analysis},
  volume = {79},
  pages = {102465},
  year = {2022},
  publisher = {Elsevier},
  isbn = {1361-8415}
}

@article{rokach2010ensemblebased,
  author = {Rokach, Lior},
  title = {Ensemble-Based Classifiers},
  journal = {Artificial intelligence review},
  volume = {33},
  number = {1},
  pages = {1--39},
  year = {2010},
  publisher = {Springer},
  isbn = {0269-2821}
}

@inproceedings{ronneberger2015unet,
  author = {Ronneberger, Olaf and Fischer, Philipp and Brox, Thomas},
  title = {U-Net: Convolutional Networks for Biomedical Image Segmentation},
  pages = {234--241},
  year = {2015},
  booktitle = {International Conference on Medical Image Computing and Computer-Assisted Intervention},
  publisher = {Springer}
}

@incollection{salgado-garcia2024efficient,
  author = {{Salgado-Garcia}, Raul J. and {Vila-Blanco}, Nicol{\'a}s and Carreira, Mar{\'i}a J. and {Nu{\~n}ez-Garcia}, Marta},
  title = {Efficient Multi-Modal Whole Heart Segmentation via Cascaded U-Net: A Practical Solution for Clinical Settings},
  pages = {158--167},
  year = {2024},
  booktitle = {MICCAI Challenge on Comprehensive Analysis and Computing of Real-World Medical Images},
  publisher = {Springer}
}

@article{schulman2017proximal,
  author = {Schulman, John and Wolski, Filip and Dhariwal, Prafulla and Radford, Alec and Klimov, Oleg},
  title = {Proximal Policy Optimization Algorithms},
  journal = {arXiv preprint arXiv:1707.06347},
  year = {2017},
  eprint = {1707.06347},
  archiveprefix = {arXiv}
}

@article{shaker2024unetr,
  author = {Shaker, Abdelrahman and Maaz, Muhammad and Rasheed, Hanoona and Khan, Salman and Yang, Ming-Hsuan and Khan, Fahad Shahbaz},
  title = {UNETR++: Delving into Efficient and Accurate 3D Medical Image Segmentation},
  journal = {IEEE Transactions on Medical Imaging},
  volume = {43},
  number = {9},
  pages = {3377--3390},
  year = {2024},
  publisher = {IEEE},
  isbn = {0278-0062}
}

@inproceedings{strudel2021segmenter,
  author = {Strudel, Robin and Garcia, Ricardo and Laptev, Ivan and Schmid, Cordelia},
  title = {Segmenter: Transformer for Semantic Segmentation},
  pages = {7262--7272},
  year = {2021},
  booktitle = {Proceedings of the IEEE/CVF International Conference on Computer Vision}
}

@article{sun2024gasaunet,
  author = {Sun, Chengkun and Terry, Russell Stevens and Bian, Jiang and Xu, Jie},
  title = {GASA-UNet: Global Axial Self-Attention U-Net for 3D Medical Image Segmentation},
  journal = {arXiv preprint arXiv:2409.13146},
  year = {2024},
  eprint = {2409.13146},
  archiveprefix = {arXiv}
}

@inproceedings{tang2022selfsupervised,
  author = {Tang, Yucheng and Yang, Dong and Li, Wenqi and Roth, Holger R. and Landman, Bennett and {et al.}},
  title = {Self-Supervised Pre-Training of Swin Transformers for 3d Medical Image Analysis},
  pages = {20730--20740},
  year = {2022},
  booktitle = {Proceedings of the IEEE/CVF Conference on Computer Vision and Pattern Recognition}
}

@article{valsangiacomobuechel2012imaging,
  author = {Valsangiacomo Buechel, Emanuela R. and Mertens, Luc L.},
  title = {Imaging the Right Heart: The Use of Integrated Multimodality Imaging},
  journal = {European heart journal},
  volume = {33},
  number = {8},
  pages = {949--960},
  year = {2012},
  publisher = {Oxford University Press},
  isbn = {1522-9645}
}

@article{wan2025srpo,
  author = {Wan, Zhongwei and Dou, Zhihao and Liu, Che and Zhang, Yu and Cui, Dongfei and {et al.}},
  title = {Srpo: Enhancing Multimodal Llm Reasoning via Reflection-Aware Reinforcement Learning},
  journal = {arXiv preprint arXiv:2506.01713},
  year = {2025},
  eprint = {2506.01713},
  archiveprefix = {arXiv}
}

@article{wang2022cycmis,
  author = {Wang, Runze and Zheng, Guoyan},
  title = {CyCMIS: Cycle-Consistent Cross-Domain Medical Image Segmentation via Diverse Image Augmentation},
  journal = {Medical Image Analysis},
  volume = {76},
  pages = {102328},
  year = {2022},
  publisher = {Elsevier},
  isbn = {1361-8415}
}

@inproceedings{wang2022mixed,
  author = {Wang, Hongyi and Xie, Shiao and Lin, Lanfen and Iwamoto, Yutaro and Han, Xian-Hua and Chen, Yen-Wei and Tong, Ruofeng},
  title = {Mixed Transformer U-Net for Medical Image Segmentation},
  pages = {2390--2394},
  year = {2022},
  booktitle = {ICASSP 2022-2022 IEEE International Conference on Acoustics, Speech and Signal Processing (ICASSP)},
  publisher = {IEEE},
  isbn = {1-6654-0540-6}
}

@inproceedings{worrall2018cubenet,
  author = {Worrall, Daniel and Brostow, Gabriel},
  title = {Cubenet: Equivariance to 3d Rotation and Translation},
  pages = {567--584},
  year = {2018},
  booktitle = {Proceedings of the European Conference on Computer Vision (ECCV)}
}

@article{wu2023dformer,
  author = {Wu, Yixuan and Liao, Kuanlun and Chen, Jintai and Wang, Jinhong and Chen, Danny Z. and Gao, Honghao and Wu, Jian},
  title = {D-Former: A u-Shaped Dilated Transformer for 3d Medical Image Segmentation},
  journal = {Neural Computing and Applications},
  volume = {35},
  number = {2},
  pages = {1931--1944},
  year = {2023},
  publisher = {Springer},
  isbn = {0941-0643}
}

@article{xu2025hrmedseg,
  author = {Xu, Qing and Lou, Zhenye and Li, Chenxin and He, Xiangjian and Qu, Rong and {et al.}},
  title = {HRMedSeg: Unlocking High-Resolution Medical Image Segmentation via Memory-Efficient Attention Modeling},
  journal = {arXiv preprint arXiv:2504.06205},
  year = {2025},
  eprint = {2504.06205},
  archiveprefix = {arXiv}
}

@inproceedings{yao2024cmragent,
  author = {Yao, Gongxin and Xuan, Yixin and Li, Xinyang and Pan, Yu},
  title = {CMR-Agent: Learning a Cross-Modal Agent for Iterative Image-to-Point Cloud Registration},
  pages = {13458--13465},
  year = {2024},
  booktitle = {2024 IEEE/RSJ International Conference on Intelligent Robots and Systems (IROS)},
  publisher = {IEEE},
  isbn = {979-8-3503-7770-5}
}

@incollection{yu2020dual,
  author = {Yu, Hong and Zha, Sen and Huangfu, Yubin and Chen, Chen and Ding, Meng and Li, Jiangyun},
  title = {Dual Attention U-Net for Multi-Sequence Cardiac MR Images Segmentation},
  pages = {118--127},
  year = {2020},
  booktitle = {Myocardial Pathology Segmentation Combining Multi-Sequence CMR Challenge},
  publisher = {Springer}
}

@incollection{zhang2020fully,
  author = {Zhang, Xiaoran and Noga, Michelle and Punithakumar, Kumaradevan},
  title = {Fully Automated Deep Learning Based Segmentation of Normal, Infarcted and Edema Regions from Multiple Cardiac MRI Sequences},
  pages = {82--91},
  year = {2020},
  booktitle = {Myocardial Pathology Segmentation Combining Multi-Sequence CMR Challenge},
  publisher = {Springer}
}

@incollection{zhao2020stacked,
  author = {Zhao, Zhou and Boutry, Nicolas and Puybareau, {\'E}lodie},
  title = {Stacked and Parallel U-Nets with Multi-Output for Myocardial Pathology Segmentation},
  pages = {138--145},
  year = {2020},
  booktitle = {Myocardial Pathology Segmentation Combining Multi-Sequence CMR Challenge},
  publisher = {Springer}
}

@article{zhou2023nnformer,
  author = {Zhou, Hong-Yu and Guo, Jiansen and Zhang, Yinghao and Han, Xiaoguang and Yu, Lequan and Wang, Liansheng and Yu, Yizhou},
  title = {Nnformer: Volumetric Medical Image Segmentation via a 3d Transformer},
  journal = {IEEE transactions on image processing},
  volume = {32},
  pages = {4036--4045},
  year = {2023},
  publisher = {IEEE},
  isbn = {1057-7149}
}

@article{zhuang2018multivariate,
  author = {Zhuang, Xiahai},
  title = {Multivariate Mixture Model for Myocardial Segmentation Combining Multi-Source Images},
  journal = {IEEE transactions on pattern analysis and machine intelligence},
  volume = {41},
  number = {12},
  pages = {2933--2946},
  year = {2018},
  publisher = {IEEE},
  isbn = {0162-8828}
}

@article{zhuang2019evaluation,
  author = {Zhuang, Xiahai and Li, Lei and Payer, Christian and {\v S}tern, Darko and Urschler, Martin and {et al.}},
  title = {Evaluation of Algorithms for Multi-Modality Whole Heart Segmentation: An Open-Access Grand Challenge},
  journal = {Medical image analysis},
  volume = {58},
  pages = {101537},
  year = {2019},
  publisher = {Elsevier},
  isbn = {1361-8415}
}

@book{zhuang2020myocardial,
  author = {Zhuang, Xiahai and Li, Lei},
  title = {Myocardial Pathology Segmentation Combining Multi-Sequence Cardiac Magnetic Resonance Images: First Challenge, MyoPS 2020, Held in Conjunction with MICCAI 2020, Lima, Peru, October 4, 2020, Proceedings},
  volume = {12554},
  year = {2020},
  publisher = {Springer Nature},
  isbn = {3-030-65651-9}
}

@inproceedings{zhuang2016multivariate,
  title={Multivariate mixture model for cardiac segmentation from multi-sequence MRI},
  author={Zhuang, Xiahai},
  booktitle={International Conference on Medical Image Computing and Computer-Assisted Intervention},
  pages={581--588},
  year={2016},
  organization={Springer}
}

@article{yan2021novel,
  title={A novel framework with weighted decision map based on convolutional neural network for cardiac MR segmentation},
  author={yan Li, Fei and Li, Weisheng and Gao, Xinbo and Xiao, Bin},
  journal={IEEE Journal of Biomedical and Health Informatics},
  volume={26},
  number={5},
  pages={2228--2239},
  year={2021},
  publisher={IEEE}
}


\end{document}